\documentclass[cup7a]{cupbook}
\usepackage{graphicx}
\usepackage{epsfig}
\usepackage{natbib}

\title{\it Chemical Evolution}
      
\author[Francesca Matteucci]{Francesca Matteucci\\ Department of Astronomy\\
University of Trieste\\ and Osservatorio Astronomico di Trieste (INAF)\\Via G.B. Tiepolo, 11, 34124 Trieste\\Italy\\ (matteucci@ts.astro.it)}

\begin{document}

\pagenumbering{roman}
\maketitle
\tableofcontents
\cleardoublepage
\pagenumbering{arabic}

\chapter[Chemical Evolution]{\it Chemical Evolution}

\section{Lecture I: basic assumptions and equations of chemical evolution}

To build galaxy chemical evolution models one needs to elucidate a number of
hypotheses and make assumptions on the basic ingredients.

\subsection{The basic ingredients}
\begin{itemize}

\item  INITIAL CONDITIONS: whether the mass of gas out of which stars will form is all present initially or it will be accreted later on. The chemical
composition of the initial gas (primordial or already enriched by a 
pregalactic stellar generation).

\item THE BIRTHRATE FUNCTION:
\begin{equation}
B(M,t)=\psi(t) \varphi(M)
\end{equation}
where:
\begin{equation}
\psi(t)=SFR
\end{equation}
is the star formation rate (SFR) and:
\begin{equation}
\varphi(M)=IMF
\end{equation}
is the initial mass function (IMF).

\item  STELLAR EVOLUTION AND NUCLEOSYNTHESIS: stellar yields, yields per stellar 
generation

\item 
SUPPLEMENTARY PARAMETERS : infall, outflow, radial flows.
\end{itemize}

\subsection{The Star Formation Rate}
Here we will summarize the most common parametrizations for the SFR in 
galaxies, as adopted by chemical evolution models:

\begin{itemize}
\item Constant in space and time and equal to the estimated present time SFR.
For example, for the local disk, the present time SFR is SFR=2-5$M_{\odot}pc^{-2} Gyr^{-1}$
(Boissier\& Prantzos, 1999).

\item Exponentially decreasing:
\begin{equation}
SFR= \nu e^{-t/ \tau_*}
\end{equation}
with $\tau_* = 5-15$ Gyr (Tosi, 1988). The quantity $\nu$is a parameter that 
we call efficiency of SF since it represents the SFR per unit mass of gas and 
is expressed in $Gyr^{-1}$.

\item The most used SFR is the Schmidt (1959) law, which assumes a dependence 
on the gas density, in particular:

\begin{equation}
SFR = \nu \sigma_{gas}^{k}
\end{equation}

where $k=1.4 \pm 0.15$, as suggested by a study of Kennicutt (1998) of local 
star forming galaxies.

\item Some variations of the Schmidt law with a dependence also on the total mass have been suggested for example by Dopita \& Ryder (1994). This  formulation
takes into account the feedback mechanism acting between supernovae ( SNe) and stellar winds injecting energy into the interstellar medium (ISM) and the galactic potential well. In other words, the SF process is regulated by the fact that in a region of recent star formation the gas is too hot to form stars and it is easily removed from that region. Before new stars could form the gas needs to cool and collapse back into the star forming region and this process 
depends on the potential well and therefore on the total mass density:

\begin{equation}
SFR= \nu \sigma_{tot}^{k_1} \sigma_{gas}^{k_2}
\end{equation}

with $k_1=0.5$ and $k_2=1.5$.

\item Kennicutt (1998) also suggested, as an alternative to the Schmidt law 
to fit the data, the following relation:

\begin{equation}
SFR= 0.017 \Omega_{gas} \sigma_{gas}\propto R^{-1} \sigma_{gas}
\end{equation}

with $\Omega_{gas}$ being the angular rotation speed of gas.

\item Finally a SFR induced by spiral density waves was suggested by
Wyse \& Silk (1989):

\begin{equation}
SFR=\nu V(R)R^{-1} \sigma_{gas}^{1.5}
\end{equation}
with R being the galactocentric distance and $V(R)$ the gas rotation velocity.

\end{itemize}

\subsection{The Initial Mass Function}
The IMF is a probability function describing the distribution of stars as a function of mass. The present day mass function is 
derived for the stars in the solar vicinity by counting the Main Sequence stars as a function of magnitude and then applying the 
mass-luminosty relation, holding for Main Sequence stars, to derive the distribution of stars as a function of mass. In order to 
derive the IMF one has then to make assumptions on the past history of SF.

The derived IMF is normally approximated by a power law:
\begin{equation}
\varphi(M)dM = aM^{-(1+x)} dM
\end{equation}
where $\varphi(M)$ is the number of stars with masses in the interval M, M+dM.

Salpeter (1955) proposed a one-slope IMF ($x=1.35$) valid for stars 
with $M> 10M_{\odot}$. 
Multi-slope ($x_1$, $x_2$, ..) IMFs have been suggested later on always for 
the solar vicinity (Scalo 1986,1998; Kroupa et al. 1993; Chabrier 2003).
The IMF is generally normalized as:
\begin{equation}
a\int^{100}_{0.1}{M \varphi(M)dM}=1
\end{equation}
where $a$ is the normalization constant and the assumed interval 
of integration is $0.1-100M_{\odot}$.

The IMF is generally considered constant in space and time with some 
exceptions such as the IMF suggested by Larson (1998) with:

\begin{equation}
x=1.35(1 + m/m_1)^{-1}
\end{equation}
where $m_1$ is variable typical mass and is associated to the Jeans mass. 
This IMF predicts then that $m_1$ is a decreasing function of time.

\subsection{The Infall Rate}

For the rate of gas accretion there are in the literature several parametrizations:
\begin{itemize}
\item The infall rate is constant in space and time and equal to the present 
time infall rate as measured in the Galaxy ($ \sim 1.0 M_{\odot} yr^{-1}$).

\item The infall rate is variable in space and time, and the most common assumption is an
exponential law (Chiosi 1980; Lacey \& Fall 1985):

\begin{equation}
IR= A(R) e^{-t/ \tau(R)}
\end{equation}
with $\tau(R)$ constant or varying with the galactocentric distance. 
The parameter $A(R)$ is derived by fitting the
present day total surface mass density, $\sigma_{tot}(t_G)$, 
at any specific galactocentric radius $R$.

\item For the formation of the Milky Way two episodes of infall have been
suggested (Chiappini et al. 1997), where during the first infall episode the stellar halo forms whereas during the 
second infall episode the disk forms. This particular infall law gives a good representation of the formation of the Milky Way.
The proposed two-infall law is: 
 
\begin{equation}
IR= A(R) e^{-t/ \tau_{H}(R)}+ B(R) e^{-(t-t_{max})/ \tau_{D}(R)}
\end{equation}

where $\tau_H(R)$ is the timescale for the formation of the halo which can be costant or vary with galactocentric distance.
The quantity $\tau_D(R)$ is the timescale for the formation of the disk and is a function of the galactocentric distance; in most of the models it is assumed to increase with $R$ (e.g. Matteucci \& Fran\c cois, 1989).

\item More recently, Prantzos (2003) suggested a gaussian law with a peak at 0.1 Gyr and a FWHM of 0.04 Gyr for the formation
of the stellar halo.

\end{itemize}

\subsection{The Outflow Rate}

The so-called galactic winds occur when the thermal energy of the gas in galaxies exceeds its
potential energy. Generally, gas outflows are called winds when the gas is lost forever from the galaxy. Only detailed dynamical simulations can suggest whether there is a wind or just an outflow of gas which will soon or later fall back again into the galaxy.
In chemical evolution models galactic winds can be sudden or continuous. If they are sudden, the mass is assumed to be lost in a very short interval of time and the galaxy is devoided from all the gas; 
if they are continuous, one has to assume the rate of gas loss.
Generally, in chemical evolution models (Bradamante et al. 1998) and also in cosmological simulations (Springel  \& Hernquist, 
2003) it is assumed that the rate of gas loss is several times the SFR:

\begin{equation}
W=- \lambda SFR
\end{equation}

where $\lambda$ is a free parameter with the meaning of wind efficiency. 
This particular formulation for the galactic wind rate is confirmed by 
observational findings (see Martin, 1999).

\subsection{Stellar evolution and nucleosynthesis: the stellar yields}

Here we summarize the various contribution to the element production by stars of all masses.
\begin{itemize}
\item Brown Dwarfs ($M < M_L$, $M_{L}=0.08-0.09M_{\odot}$) are
objects which never ignite H and their lifetimes are larger than the age 
of the Universe. They are contributing to lock up mass.

\item Low mass stars ($0.5 \le M/M_{\odot} \le M_{HeF}$) 
(1.85-2.2$M_{\odot}$) 
ignite He explosively but without destroying themselves and then
become C-O  white dwarfs (WD).
If $M < 0.5 M_{\odot}$
they become He WDs. Their lifetimes range from several $10^{9}$ years 
up to several Hubble times!

\item Intermediate mass stars 
($M_{HeF} \le M/M_{\odot} \le M_{up}$) ignite He quiescently.
The mass $M_{up}$ is the limiting mass for the formation of a C-O 
degenerate core
and is in the range 5-9$M_{\odot}$, depending on stellar 
evolution calculations.
Lifetimes are from several $10^{7}$ to  $10^{9}$ years.
They die as C-O WDs if not in binary systems.
If in binary systems they can give rise to cataclysmic variables such
as novae and Type Ia SNe.

\item  Massive stars ($M >M_{up}$).
We distinguish here several cases: \par
-$M_{up} \le M/M_{\odot} \le 10-12$. Stars with Main Sequence masses in 
this range end up as electron-capture SNe 
leaving
neutron stars as remnants. These SNe will appear as Type II SNe which show H in their spectra.\par
-$10-12 \le M/M_{\odot} \le M_{WR}$, (with $M_{WR} \sim 20-40 M_{\odot}$ being
the limiting mass for the formation of a Wolf-Rayet (WR) star). 
Stars in this mass range
end their life as core-collapse SNe (Type II) leaving a
neutron star or a black hole as remnants. \par
-$M_{WR} \le M/M_{\odot} \le 100$. Stars in this mass range are probably 
exploding as Type Ib/c SNe which do not show H in their spectra. Their
lifetimes are of the order of $\sim 10^{6}$ years.

\item Very Massive Stars ($M > 100 M_{\odot}$), 
they should explode by means of  instability due to ``pair creation'' 
and they are called {\it pair-creation} SNe. 
In fact, at $T \sim 2 \cdot 10^{9}$ K a large portion of the gravitational 
energy 
goes into creation of pairs $(e^{+}, e^{-})$, the star 
becomes unstable and explodes. They leave no remnants and their 
lifetimes are $< 10^{6}$ years. Probably these very massive stars formed only when the metal content was almost zero (Population III stars, Schneider et al. 2004). 

\end{itemize}

All the elements with mass number $A$ from 12 to 60 have 
been formed in stars during
the quiescent burnings.
Stars transform H into He and then He into 
heaviers until the 
Fe-peak elements, where the binding energy per nucleon reaches a maximum 
and the nuclear fusion reactions stop.

H is transformed into He through the proton-proton 
chain or the 
CNO-cycle, then $^{4}He$ is transformed into $^{12}C$ through the 
triple- $\alpha$
reaction.

Elements heavier than $^{12}C$ are then produced by synthesis 
of $\alpha$-particles. They are called $\alpha$-elements 
(O, Ne, Mg, Si and others).

The last main burning in stars is the $^{28}Si$ -burning which produces
$^{56}Ni$ which then decays into $^{56}Co$ and $^{56}Fe$.
Si-burning can be quiescent or explosive (depending on the temperature).

Explosive nucleosynthesis 
occurring during SN explosions 
mainly produces Fe-peak elements. Elements
originating from s- and r-processes (with A$> 60$ up to Th and U)
are formed by means of slow or rapid (relative to the $\beta$- decay)
neutron capture by Fe seed nuclei;
s-processing occurs during quiescent He-burning 
whereas r-processing occurs during SN explosions. 

\subsection{Type Ia SN Progenitors}

The Type Ia SNe, which do not show H in their spectra, are believed 
to originate from WDs in binary 
systems and to be the major producers of Fe in the Universe.
The model proposed are basically two:

\begin{itemize}
\item {\bf Single Degenerate Scenario (SDS)}, with a WD plus a 
Main Sequence or Red Giant star, as  originally suggested by 
Whelan and Iben (1973). The explosion (C-deflagration) occurs when the 
C-O WD reaches the Chandrasekhar mass, $M_{Ch}=\sim 1.44 M_{\odot}$,  
after accreting material from thecompanion. In this model the clock to the 
explosion is given by the lifetime of the companion of the WD (namely the 
less massive star in the system). It is interesting to define the minimum 
timescale for the explosion which is given by the lifetime of a $8 M_{\odot}$ 
star, namely $t_{SNIa_{min}}$=0.03 Gyr 
(Greggio and Renzini 1983). Recent observations in radio-galaxies by Mannucci 
et al. (2005;2006) seem to confirm the existence of such prompt Type Ia SNe.

\item {\bf Double Degenerate Scenario (DDS)}, where 
the merging of two C-O WDs of mass $\sim 0.7 M_{\odot}$, 
due to loss of angular momentum as a consequence of 
gravitational wave radiation, 
produces C-deflagration (Iben
and Tutukov 1984). In this case the clock to the explosion is given by the lifetime of the secondary star, as above, plus the gravitational time delay, namely the time necessary for the two WDs to merge. The minimum time for the explosion is
$t_{SNIa_{min}}=0.03 +\Delta t_{grav}$=0.04 Gyr 
(see Tornamb\`e 1989).
\end{itemize}

Some variations of the above scenarios have been proposed such as the model
by Hachisu et al. (1996; 1999), which is based on the single degenerate 
scenario where a wind from the WD is considered. Such a wind stabilizes the accretion from the companion and introduces a metallicity effect. In particular,
the wind, necessary to this model, occurs only if the systems have metallicity 
([Fe/H]$< -1.0$). This implies that the minimum time for the explosion is larger than in the previous cases. In particular,
$t_{SNIa_{min}} = 0.33 $ Gyr, which is the lifetime of the more massive 
secondary considered ($2.3 M_{\odot}$)
plus the metallicity delay which depends on the assumed chemical evolution model.

\subsection{Yields per Stellar Generation}

Under the assumption of Instantaneous Recycling Approximation  (IRA)
which states that all stars more massive than $1M_{\odot}$ die immediately, whereas all stars with masses lower than $1M_{\odot}$ live forever, one can define the yield per stellar generation (Tinsley, 1980);
\begin{equation}
y_{i}={1 \over 1-R} \int^{\infty}_{1}{m p_{im} \varphi(m) dm} 
\end{equation}

where $p_{im}$ is the stellar yield of the element $i$, namely the newly 
formed and ejected element $i$ by a star of mass $m$.

The quantity $R$ is the so-called Returned Fraction:
\begin{equation}
R=\int^{\infty}_{1}{(m-M_{rem}) \varphi(m) dm} 
\end{equation}
and is the total mass of gas restored into the ISM by an entire stellar 
generation.

\subsection{Analytical models}
The {\it Simple Model} for
the chemical evolution of the solar neighbourhood is the simplest approach to 
model chemical evolution. The solar neighbourhood is assumed to be a cylinder of 1 Kpc radius centered around the Sun.

The basic assumptions of the Simple Model are:\par

- the system is one-zone and closed, no inflows 
or outflows with the total mass present since the beginning,

- the initial gas is primordial (no metals),

- instantaneous recycling approximation holds, 
 
- the IMF, $\varphi(m)$, is assumed to be constant in time,

- the gas is well mixed at any time (IMA)\par

The Simple Model fails in describing the evolution of the Milky Way 
(G-dwarf metallicity distribution, elements produced on long timescales and 
abundance ratios) and the reason is that at least
two of the above assumptions are manifestly wrong, epecially if one intends to model the evolution of the abundance of elements produced on long timescales, such as Fe. In particular the 
assumptions of the closed boxiness and the IRA. 

However, it is interesting 
to know the solution of the Simple Model and its implications.
Be $X_i$ the abundance by mass of an element $i$.

If
$X_i<<1$, which is generally true for metals, we obtain the solution
of the Simple Model. This solution is obtained analytically by ignoring the stellar lifetimes:

\begin{equation}
X_i= y_{i} ln({ 1 \over G}) 
\end{equation}
where $\mu =M_{gas}/M_{tot}$
and  $y_i$ is the yield per stellar generation, as defined above, otherwise called {\it effective yield}.
In particular, the effective yield is defined as:
\begin{equation}
y_{i_{eff}}={X_i \over ln(1/G)} 
\end{equation}
namely the yield that the system would have if behaving as the simple 
closed-box model.
This means that if $y_{i_{eff}} > y_{i}$, then the actual system has 
attained a higher abundance for the element $i$ at a given gas fraction G.
Generally, in the IRA,
we can assume:

\begin{equation}
{X_i \over X_j}= {y_i \over y_j}
\end{equation}

which means that the ratio of two element abundances are always equal to the 
ratio of their yields.
This is no more  true when IRA is relaxed. In fact, relaxing IRA is 
necessary  to study in detail the evolution of the abundances of single 
elements.

One can obtain analytical solutions also in presence of infall and/or outflow but the necessary condition is to assume IRA.
Matteucci \& Chiosi (1983) found solutions for models with outflow and infall 
and Matteucci (2001) found it for a model with infall and outflow acting at 
the same time.
The main assumption in the model with outflow but no infall
is that the outflow rate is:

\begin{equation}
W(t)=\lambda (1-R) \psi(t)
\end{equation}

where $\lambda \ge 0$ is the wind parameter.

The solution of this model is:

\begin{equation}
X_i = {y_{i} \over (1+ \lambda)} ln[(1+ \lambda) G^{-1} - \lambda]
\end{equation}

for $\lambda = 0$ the equation becomes the one of the Simple Model (1.17).

The solution of the equation of metals for a model without wind but with
a primordial infalling material ($X_{A_{i}}=0$) at a rate: 

\begin{equation}
A(t)=\Lambda (1-R) \psi(t)
\end{equation}

and $\Lambda \ne 1$  is :
\begin{equation}
X_i= {y_i \over \Lambda}[1-(\Lambda-(\Lambda-1)G^{-1})^
{-\Lambda/(1-\Lambda)}]
\end{equation}
For $\Lambda=1$ one obtains the well known case of {\it extreme infall} studied by Larson (1972) whose solution is:
\begin{equation}
X_i=y_i[1-e^{-(G^{-1} -1)}]
\end{equation}
This extreme infall solution shows that when $G \rightarrow 0$ then $X_i \rightarrow y_i$.

\subsection{Numerical Models}

Numerical models relax IRA and close boxiness but generally retain
the constancy of $\varphi(m)$ and the IMA.

If $G_i$ is the mass fraction of gas in the form of an element $i$, 
we can write:

\begin{eqnarray}
 & & \dot G_i(t)  =  -\psi(t)X_i(t)\nonumber \\
& & + \int_{M_{L}}^{M_{Bm}}\psi(t-\tau_m)
Q_{mi}(t-\tau_m)\varphi(m)dm\nonumber \\ 
& & + A\int_{M_{Bm}}^{M_{BM}}
\phi(m)\nonumber \\
& & \cdot[\int_{\mu_{min}}
^{0.5}f(\gamma)\psi(t-\tau_{m2}) 
Q_{mi}(t-\tau_{m2})d\gamma]dm\nonumber \\ 
& & + B\int_{M_{Bm}}^
{M_{BM}}\psi(t-\tau_{m})Q_{mi}(t-\tau_m)\varphi(m)dm\nonumber \\
& & + \int_{M_{BM}}^{M_U}\psi(t-\tau_m)Q_{mi}(t-\tau_m) 
\varphi(m)dm\nonumber \\ 
& & + X_{A_{i}} A(t) - X_{i}(t) W(t)
\end{eqnarray}

where B=1-A, A=0.05-0.09. The meaning of the A parameter is the fraction in the IMF of binary systems with those specific features required to give rise to Type Ia SNe, whereas B is the fraction of all the single stars and binary systems in the same mass range of definition of the progenitors of Type Ia SNe.
The values of A indicated above are correct for the evolution of the solar vicinity where an IMF of Scalo (1986, 1989) or Kroupa et al.(1993) is adopted.
If one adopts a flatter IMF such as the  Salpeter (1955) one then A is 
different. In the above equations the contribution of Type Ia SNe is contained in the third term on the right hand side. The integral is made over a range of masses going from 3 to 16 $M_{\odot}$ which represents the total masses of binary systems able to produce Type Ia SNe in the framework of the SDS. There is also an integration over the mass distribution of binary systems; in particular, one considers the function $f(\gamma)$ where $\gamma={M_2 \over M_1 +M_2}$,
with $M_1$ and $M_2$ being the primary and secondary mass of the binary system, respectively (for more details see Matteucci \& Greggio 1986 and Matteucci 2001).The functions A(t) and W(t) are the infall and wind rate, respectively. 
Finally, the quantity $Q_{mi}$ represents the stellar yields (both processed 
and unprocessed material).

\section{Lecture II: the Milky Way and other spirals}

The Milky Way galaxy has four main stellar populations: 1) the halo stars with low metallicities (the most common metallicity indicator in stars is [Fe/H]=
$log(Fe/H)_{*}- log(Fe/H)_{\odot}$) and eccentric orbits, 2) the bulge population with a large range of metallicities and is dominated by random motions, 3) the thin disk stars with an average metallicity $<[Fe/H]>$=-0.5 dex and circular orbits, and finally 4) the thick stars which possess chemical and kinematical properties intermediate between those of the halo and those of the thin disk. The halo stars have average metallicities of $<[Fe/H]>$=-1.5 dex and a maximum metallicity of $\sim -1.0$ dex although stars with [Fe/H] as high as -0.6 dex and halo kinematics are observed. 
The average metallicity of thin disk stars is $\sim -0.6$ dex, whereas the one of Bulge stars is $\sim -0.2$ dex.

\subsection{The Galactic formation timescales}
The kinematical and chemical properties of the different Galactic stellar populations can be interpreted in terms of 
the Galaxy formation mechanism.
Eggen et al. (1962) in a cornerstone paper suggested a rapid collapse for the formation of the Galaxy lasting $\sim 3 \cdot 10^{8}$ years. This suggestion was based on a kinematical and chemical study of solar neighbourhood stars. Later on, Searle \& Zinn (1979) proposed a central collapse like the one proposed by Eggen et al. 
but also that the outer halo formed by merging of large fragments taking place over a considerable timescale $> 1$ Gyr.
More recently, Berman \& Suchov (1991) proposed the so-called hot Galaxy 
picture, with an initial strong burst of SF which inhibited  further SF for 
few Gyr while a strong Galactic wind was created.

From an historical point of view, the modelization of the Galactic chemical 
evolution has passed through different phases that I summarize in the following.

\begin{itemize} 
\item SERIAL FORMATION

The Galaxy is modeled by means of one accretion episode lasting for the entire Galactic lifetime, where halo, thick and thin disk form in sequence as a continuous process. The obvious limit of this approach 
is that it does not allow us to predict the observed overlapping in metallicity between halo and thick disk stars and between thick and thin disk stars, but it gives a fair representation of our Galaxy  
(e.g. Matteucci  \& Fran\c cois 1989).
\item PARALLEL FORMATION

In this formulation, the various Galactic components start at the same time 
and 
from the same gas but evolve at different rates (e.g. Pardi et al. 1995). It
predicts overlapping of stars belonging to the different components
but implies that the thick disk formed out of gas shed by the halo and that the thin disk formed out of gas shed by the thick disk, and this is at variance with the distribution of the stellar angular momentum per unit mass (Wyse \& Gilmore 1992), which indicates that the disk did not form out of gas shed by the halo.

\item TWO-INFALL FORMATION

In this scenario, halo and disk 
formed  out of two separate infall episodes (overlapping in metallicity 
is also predicted)
(e.g. Chiappini et al. 1997; Chang et al. 1999). The first infall 
episode lasted no more than 1-2 Gyr whereas the second, where the thin disk formed, 
lasted much longer with a timescale for the formation of the solar vicinity of 6-8 Gyr 
(Chiappini et al. 1997; Boissier\& Prantzos 1999).

\item STOCHASTIC APPROACH

Here the hypothesis is that in the early halo phases ([Fe/H] $< -3.0$ dex), 
mixing was not 
efficient and, as a consequence, one should observe in low metallicity halo 
stars the 
effects of pollution from single SNe (e.g. Tsujimoto et al. 1999; Argast et 
al. 2000; Oey 2000). 
These models predict a large spread for  [Fe/H] $< -3.0$dex  which is not 
observed, as shown 
by recent data with metallicities down to -4.0 dex 
(Cayrel et al. 2004; see later).

\end{itemize}

\subsection{The two-infall model}
The adopted SFR  (see Figure 2.1) is eq.(1.6) with different SF efficiencies for the halo and disk, 
in particular $\nu_H=2.0 Gyr^{-1}$, $\nu_D=1.0 Gyr^{-1}$, respectively. A threshold density 
($\sigma_{th}=7 M_{\odot} pc^{-2}$) for the SFR is also assumed in agreement with results from Kennicutt (1989; 1998).

\begin{figure}
\includegraphics[width=4.5in,height=3.0in]{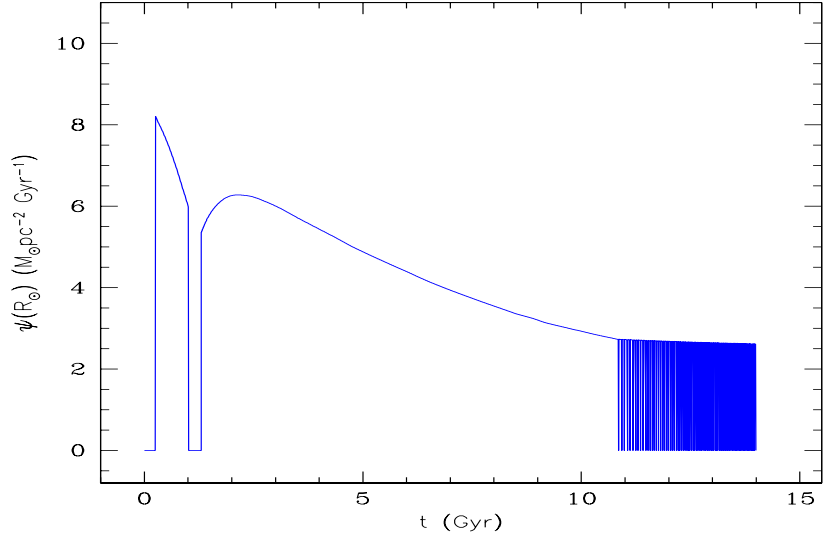}
\hfill
\caption{The predicted SFR in the solar vicinity with the two-infall model. Figure from Chiappini et al. (1997). The oscillating behaviour at late times is due to the assumed threshold density for SF. The threshold gas density is also 
responsible for the gap in the SFR seen at around 1 Gyr.}\label{fig} 
\end{figure}

In Figure 2.2 we show the predicted SN (II and Ia) rates by the two-infall model. Note that the Type Ia SN rate is calculated according to the SDS (Greggio \& Renzini, 1983; Matteucci \& Recchi, 2001). There is a  delay between the Type II SN rate and the Type Ia SN rate, and while the Type II SN rate strictly follows the SFR,  the Type Ia SN rate is smoothly increasing.

\begin{figure}
\includegraphics[width=4.5in,height=3.0in]{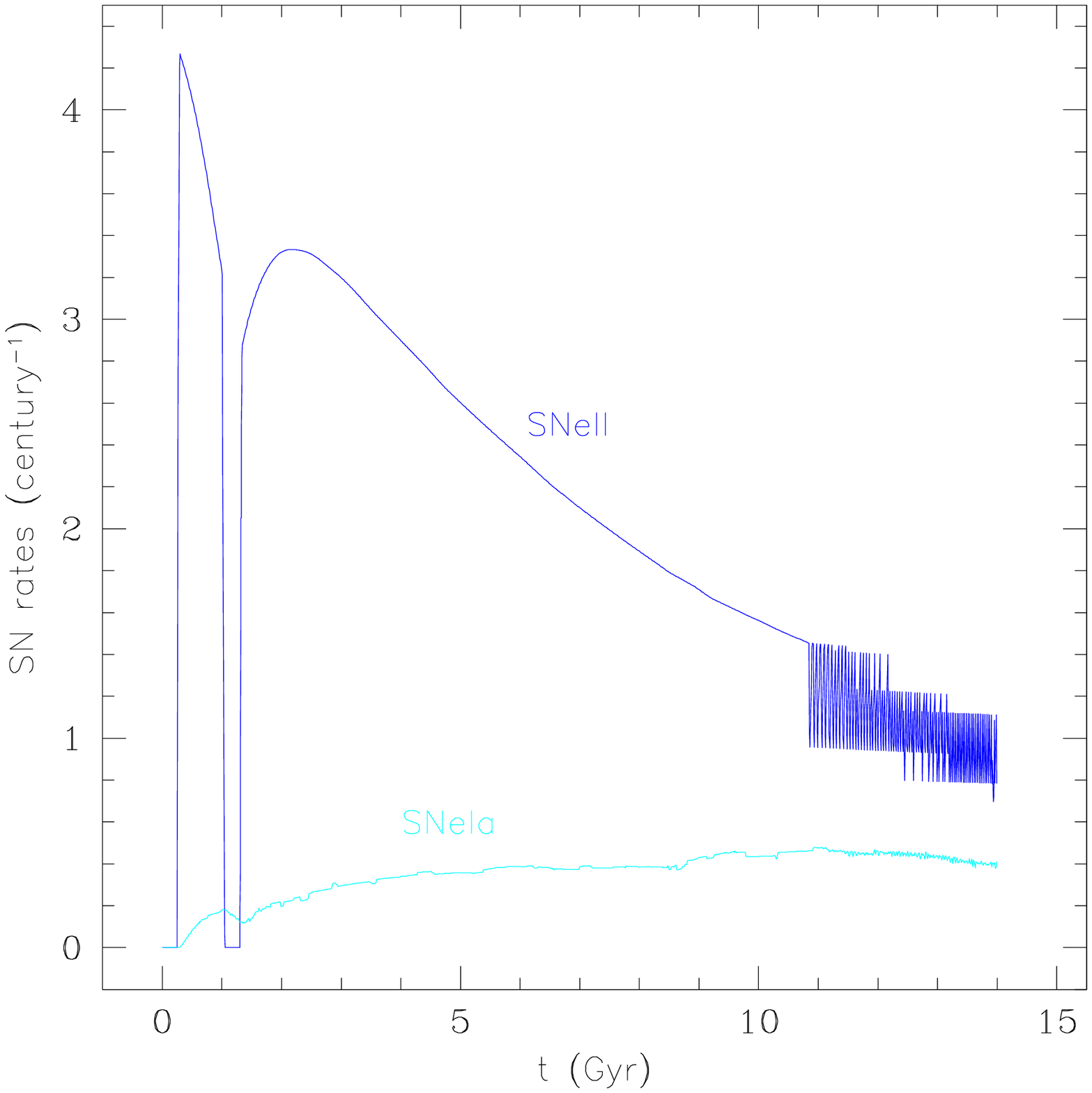}
\hfill
\caption{The predicted Type II and Ia SN rate in the solar vicinity with the two-infall model. Figure from Chiappini et al. (1997)}\label{fig} 
\end{figure}

Fran\c cois et al. (2004) compared the predictions of the two-infall 
model for the abundance ratios versus metallicity relations ([X/Fe] vs. [Fe/H]), with the very recent and very 
accurate data of the project ``First Stars'' by Cayrel et al. (2004). They adopted yields from the literature both for Type II and Type Ia SNe and noticed that while for some elements (O, Fe, Si, Ca) the yields of Woosley \& Weaver (1995) (hereafter WW95) reproduce the data fairly well, 
for the Fe-peak elements and heaviers none of the available yields give a good 
agreement. Therefore, they varied empirically the yields of these elements in order to best fit the data. In Figures 2.3 and 2.4 we show the predictions for $\alpha$-elements (O, Mg, Si, Ca, Ti, K) plus some Fe-peak elements and Zn.

\begin{figure}
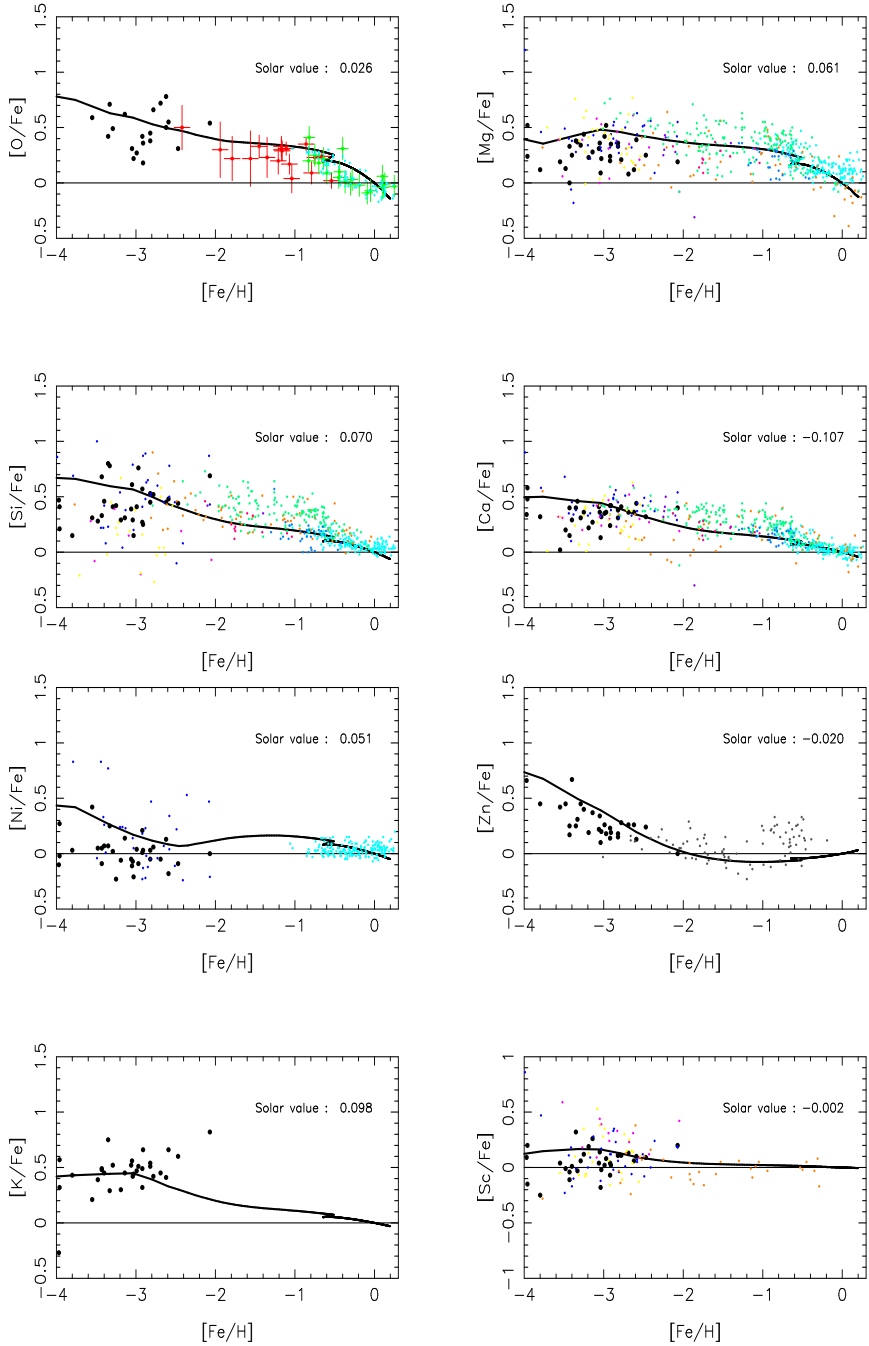

\includegraphics[width=4.5in,height=3.5in]{Fig3Matteucci.eps}
\includegraphics[width=4.5in,height=3.5in]{Fig4Matteucci.eps}
\hfill
\caption{Predicted and observed [X/Fe] vs. [Fe/H] for several $\alpha$- and Fe-peak- elements plus Zn compared with a 
compilation 
of data. In particular the black dots are the recent high resolution data from Cayrel et al. (2004). For the other data 
see references in Fran\c cois et al. (2004). The solar value indicated in the upper right part of each figure represents 
the predicted solar value for the ratio [X/Fe]. The assumed solar abundances are those of 
Grevesse \& Sauval (1998) except that for oxygen for which we take the value of Holweger (2001). }\label{fig} 
\end{figure}

In Figure 2.4 we show 
also the ratios between the yields derived empirically by Fran\c cois et al. (2004) in order to obtain the excellent fits 
shown in the 
figures, and those of WW95 for massive stars. For some elements it was necessary to change also the yields from Type Ia SNe relative 
to the reference ones which are those of  Iwamoto et al. (1999)
(hereafter I99).

\begin{figure}
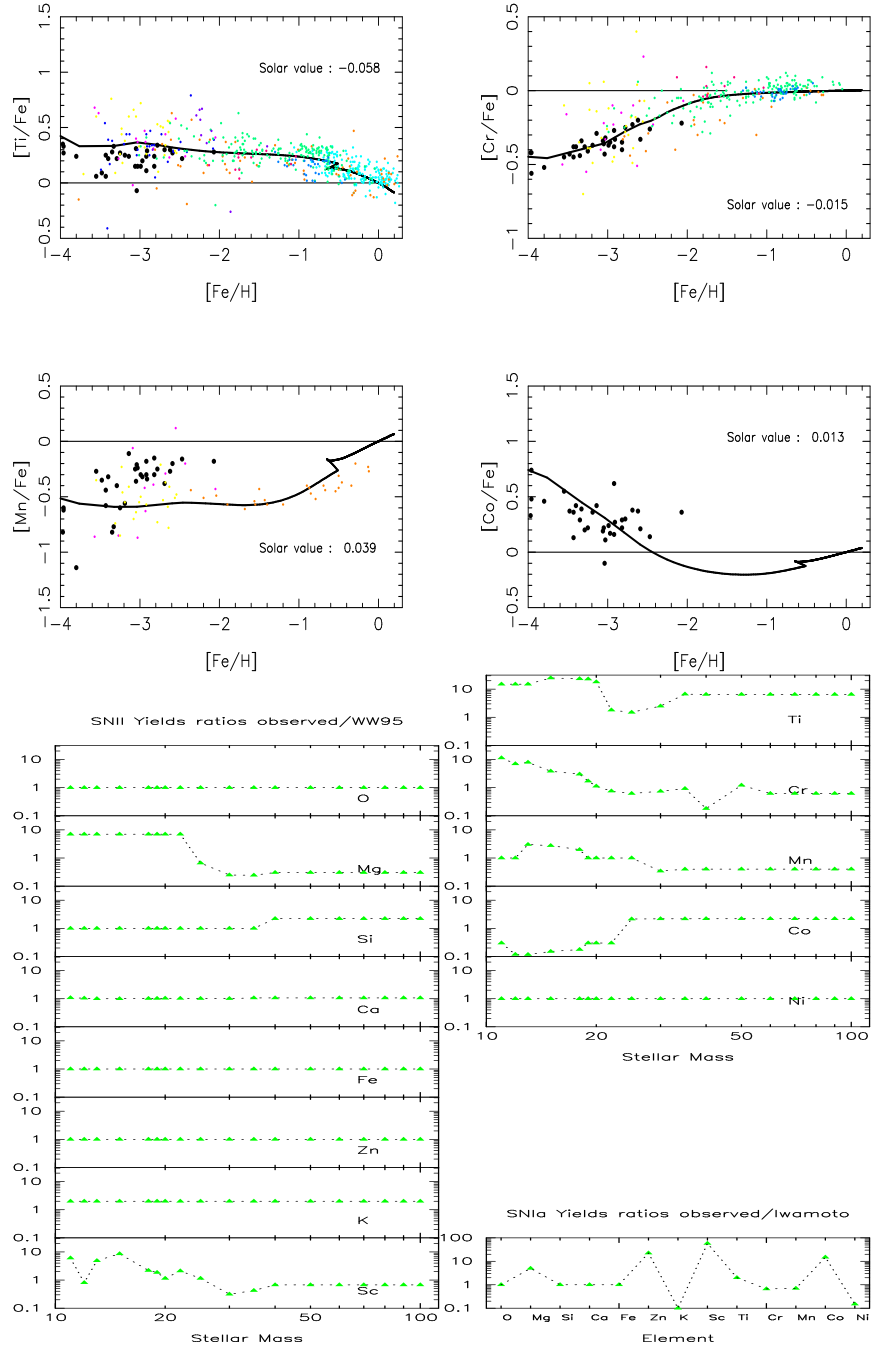

\includegraphics[width=4.5in,height=3.5in]{Fig5Matteucci.eps}
\includegraphics[width=4.5in,height=3.5in]{Fig6Matteucci.eps}
\hfill
\caption{Upper panel: predicted and observed [X/Fe] vs. [Fe/H] for several elements as in 
Figure 2.3. In the bottom part of this Figure are shown the ratios between the empirical yields and the yields by WW95 
for massive stars. Such empirical yields have been suggested by Fran\c cois et al. (2004) in order to fit at best all 
the [X/Fe] vs. [Fe/H] relations. In the small panel at the bottom right side are shown also the ratios between the 
empirical yields for Type Ia SNe and the yields by I99.}\label{fig} 
\end{figure}

In Figure 2.5 we show the predictions of chemical evolution models for $^{12}C$ and 
$^{14}N$ compared with abundance data. The behaviour of C  shows a roughly constant 
[C/Fe] as a function of [Fe/H], although C seems to  slightly increase at very 
low metallicities, indicating that the bulk of these two elements comes from 
stars with the same lifetimes. The data in these figures, especially those 
for N are old
and do not contain very metal poor stars. Newer data containing stars with 
[Fe/H] down to $\sim$ -4.0 dex (Spite et al. 2005; Israelian et al. 2004) 
indicate that the [N/Fe] ratio 
continues to be high also at low metallicities, indicating a primary origin 
for N produced in massive stars. We recall here that we define {\it primary} 
a chemical elements which is produced in the stars starting from the H and He, 
whereas we define {\it secondary} a chemical element which is formed from 
heavy elements already present in the star at its birth and not produced in 
situ. 
The model predictions shown in Figure 2.5 for C and N assume that the bulk of 
these elements is produced by low and intermediate mass stars (yields from van 
den Hoeck and Groenewegen, 1997) and that N is produced as a
partly secondary and partly primary element. 
The N production from massive stars has only a secondary origin (yields from WW95).
In Figure 2.5 we show also a model prediction where N is considered as a primary element in massive stars with the yields artificially increased. Recently, Chiappini et al. (2006) have shown that primary N produced by very metal poor fastly rotating massive stars can well reproduce the observations.

\begin{figure}
\includegraphics[width=4.5in,height=3.0in]{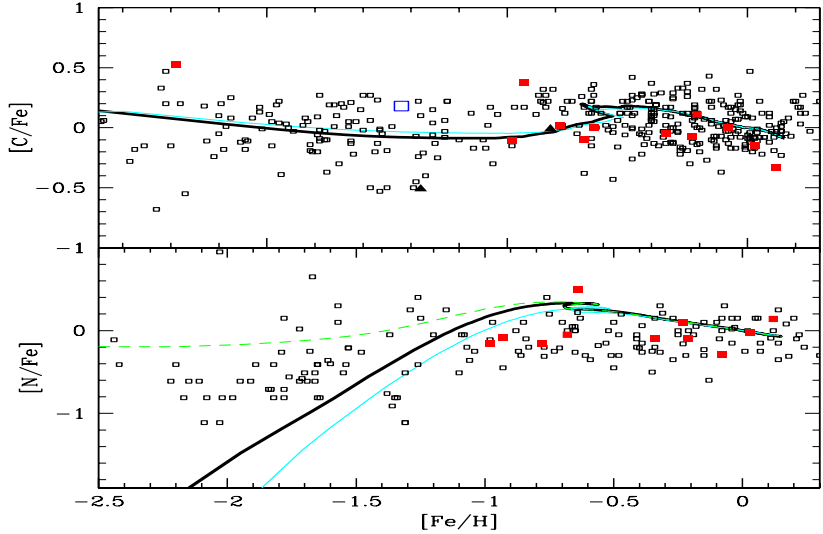}
\hfill
\caption{Upper panel: predicted and observed [C/Fe] vs. [Fe/H]. Models from Chiappini et al. (2003a). Lower panel, predicted and observed [N/Fe] vs. [Fe/H]. For references to the data see original paper.The thin and thick continuous lines in both panels represent models with standard nucleosynthesis, as described in the text, whereas the dashed line represents the predictions of a model where N in massive stars has been considered as a primary element with ``ad hoc'' stellar yields. }\label{fig} 
\end{figure}

In summary, the comparison between model predictions and abundance data 
indicate the following scenario for the formation of heavy elements:
\begin{itemize}
\item $^{12}C$ and $^{14}N$ are mainly produced in low and intermediate 
mass stars ($0.8 \le M/M_{\odot} \le 8$). 
The amounts of primary and secondary N is still uncertain and also the fraction of C produced in massive stars. Primary N from massive stars seems to be required to reproduce the N abundance in low metallicity halo stars.

\item
$\alpha$-elements originate in massive stars: the
nucleosynthesis of O is rather well understood
(there is agreement between different authors), the yields from WW95 as functions of metallicity produce an excellent agreement 
with the observations for this particular element.

\item Magnesium is generally underproduced by nucleosynthesis models. Taking the yields of WW95 as a reference,
the Mg yields should be increased in stars with masses 
$M \le 20 M_{\odot}$ 
and decreased in stars with $M  > 20 M_{\odot}$ to fit the data.  
Silicon should be slightly increased in stars with masses $M>40M_{\odot}$.

\item Fe  originates mostly in Type Ia SNe.
The Fe yields in massive stars are still uncertain, WW95 metallicity
dependent yields overestimate Fe
in  stars $< 30M_{\odot}$. For this element, it is better to adopt the yields of WW95 for solar 
metallicity.

\item Fe-peak elements: the yields of Cr, Mn should be increased
in stars of 10-20 $M_{\odot}$ relative to the yields of WW95, whereas the yield of
Co should be increased in Type Ia SNe, relative to the yields of I99, and decreased in stars in the 
range 10-20$M_{\odot}$, relative to the yields of WW95. Finally, the yield of
Ni should be decreased in Type Ia SNe.

\item The yields of Cu and Zn from Type Ia SNe should be larger, relative to 
the standard yields, as already suggested by Matteucci et al. (1993).

\end{itemize}

\subsection{Common Conclusions from MW Models}
Most of the chemical evolution models for the Milky Way existing 
in the literature conclude that:

\begin{itemize}
\item The G-dwarf metallicity distribution
can be reproduced only by assuming a slow
formation of the local disk by infall. In particular,
the time-scale for the formation of the local disk should be in the range
$\tau_d \sim 6-8$ Gyr
(Chiappini et al. 1997;
Boissier and Prantzos 1999; Chang et al. 1999; Chiappini et al. 2001;
Alib\`es et al. 2001).

\item  The relative abundance ratios [X/Fe] vs. [Fe/H], interpreted as
time-delay between Type Ia and II SNe, suggest a
timescale for the halo-thick
disk formation of 
$\tau_h \sim$ 1.5-2.0 Gyr (Matteucci and Greggio 1986;
Matteucci and Fran\c cois, 1989; 
Chiappini et al. 1997). The external halo and thick disk probably formed more slowly or have been accreted (Chiappini et al. 2001).

\item To fit abundance gradients, SFR and gas distribution along the Galactic 
thin disk we  must assume that the disk
formed {\it inside-out} (Matteucci \& Fran\c cois, 1989; 
Chiappini et al. 2001; Boissier \& Prantzos 1999; Alib\'es et al. 2001).
Radial flows can help in forming the gradients (Portinari \& Chiosi 2000) but they are probably not the main cause for them. A variable IMF along the Disk can in principle explain abundance gradients but it creates unrealistic situations: in fact, in order to reproduce the negative gradients one should assume that in the external and less metal rich parts of the Disk low mass stars form 
preferentially (see Chiappini et al. 2000 for a discussion on this point).

\item The SFR is a strongly varying function of the 
galactocentric distance (Matteucci \& Fran\c cois 1989; 
Chiappini et al, 1997,2001;
Goswami \& Prantzos 2000; Alib\'es et al. 2001).

\end{itemize}

\subsection{Abundance Gradients from Emission Lines}

There are two types of abundance determinations in HII regions:
one is based on recombination lines which should have  a weak temperature dependence of the nebula (He, C, N, O),
the other is based on collisionally excited lines where a strong dependence is intrinsic to the method 
(C, N, O, Ne, Si, S, Cl, Ar, Fe and Ni). 
This second method has predominated until now.
A direct determination of the abundance gradients from HII regions in the Galaxy from optical lines is difficult because 
of extinction, so usually the abundances for distances larger than 3 Kpc from the Sun are obtained from radio and infrared 
emission lines.

Abundance gradients can also be derived from optical emission lines in Planetary Nebulae (PNe). However, the abundances of He, C and N in PNe
are giving only information on the internal nucleosynthesis of the star. So, to derive gradients one should look at the abundances of O, S and Ne, unaffected
by stellar processes. In Figure 2.6 we show theoretical predictions of abundance gradients along the disk of the Milky Way compared with data from HII regions and B stars. The adopted model is from Chiappini et al. (2001; 2003a) and is based on an 
inside-out formation of the thin disk with the inner regions forming faster than the outer ones, in particular 
$\tau(R)=0.875R - 0.75$ Gyr. 
Note that to obtain a better fit for $^{12}C$, the yields of this element have been increased artificially relative to those
of WW95.

\begin{figure}

\includegraphics[width=4.5in,height=3.0in]{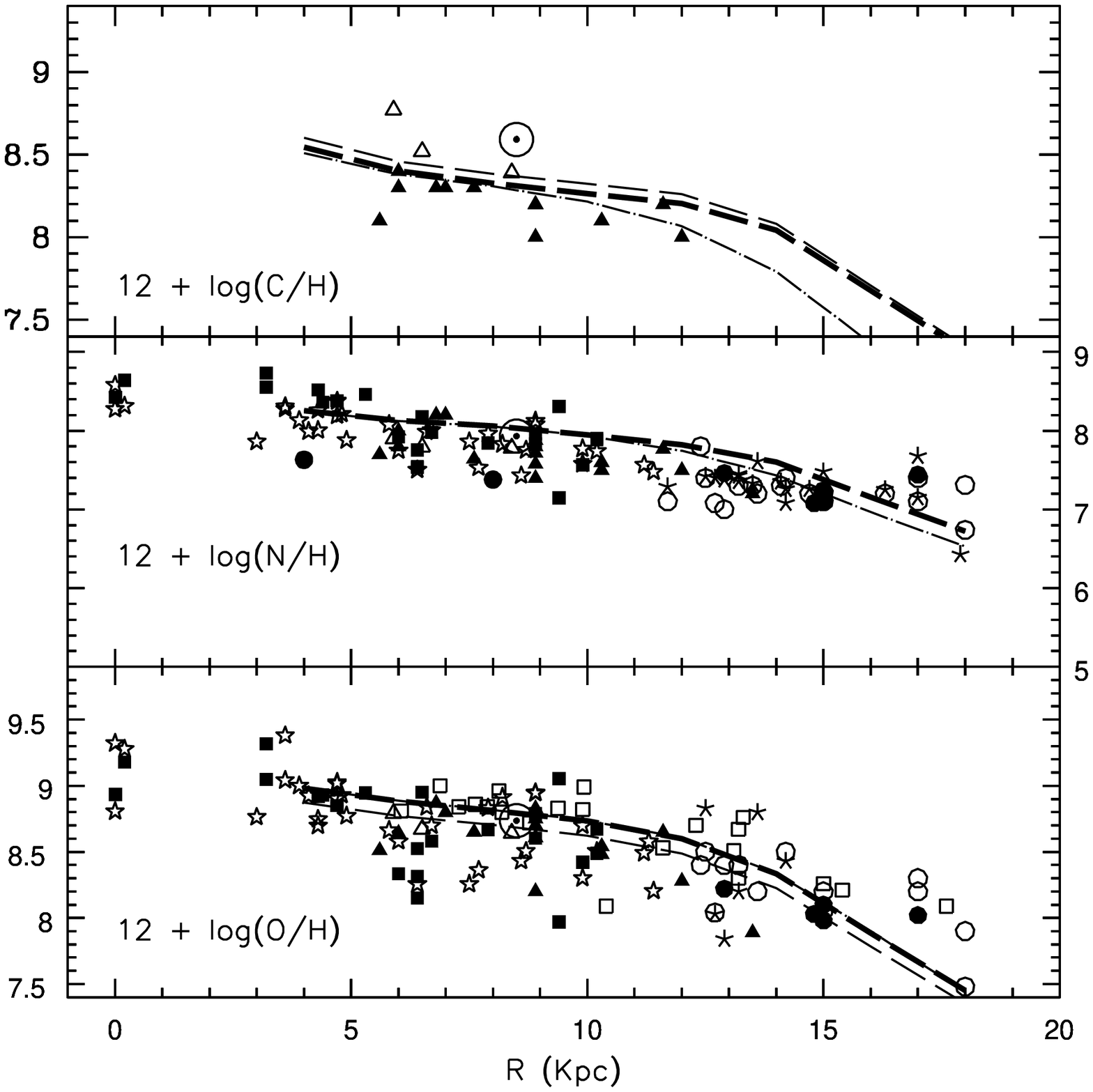}
\includegraphics[width=4.5in,height=3.0in]{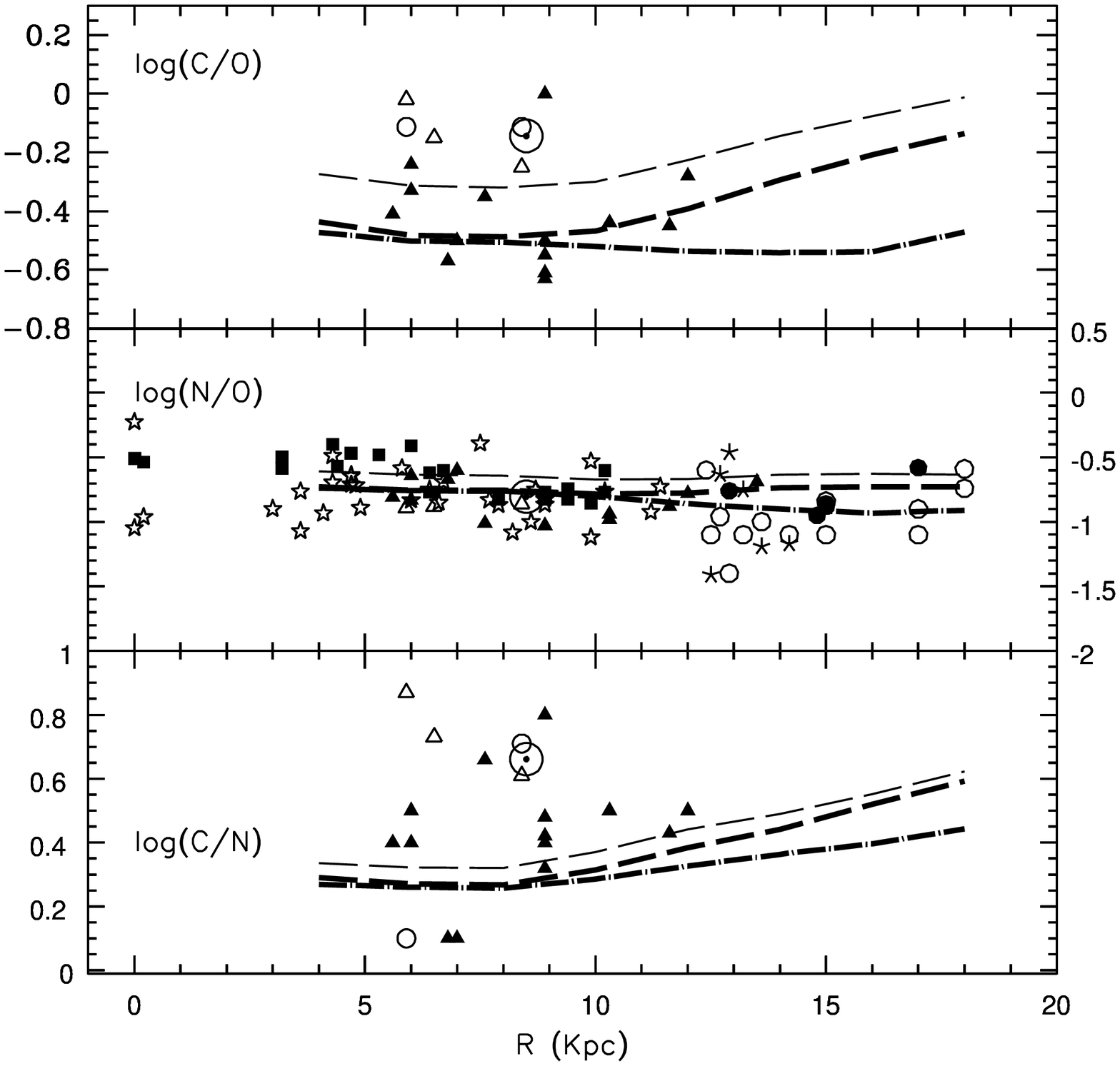}
\hfill
\caption{Upper panel: abundance gradients along the Disk of the MW. The lines are the models from Chiappini et al. (2003a): 
these models differ by the nucleosynthesis prescriptions. In particular, the dash-dotted line represents a model with 
van den Hoeck \& Groenewegen (1997, hereafter HG97) yields for low-intermediate mass stars with $\eta$ (mass loss parameter) constant and 
Thielemann et al.'s (1996) yields for massive stars, the 
long- dashed thick line has HG97 yields with variable $\eta$ and Thielemann et al. yields, the long-dashed thin line has 
HG97 yields with variable $\eta$ but WW95 yields for massive stars. It is interesting to note that in all of these models 
the yields of $^{12}C$ in stars $> 40M_{\odot}$ have been artificially increased by a factor of 3 relative to the yields 
of WW95.
Lower panel: the temporal behaviour of abundance 
gradients along the Disk as 
predicted by the best model of Chiappini et al. (2001). The upper lines in each panel represent the present time gradient, whereas the lower ones represent the gradient  a few Gyr ago. It is clear that the gradients tend to stepeen in time, a still controversial 
result.}\label{fig} 
\end{figure}

As already said, most of the models agree on the inside-out scenario for the 
Disk formation, however not all models agree on the evolution of the gradients 
with time. In fact, some models predict a flattening with time (Boissier and 
Prantzos 1998; Alib\`es et al. 2001), whereas others such as that of Chiappini 
et al. (2001) predict a steepening. The reason for the steepening is that in 
the model of Chiappini et al. is included a threshold density for SF,, which induces the SF to stop when the density decreases 
below the threshold. This effect is particularly strong in the external 
regions of the Disk, thus
contributing to a slower evolution and therefore to a steepening of the 
gradients with time, as shown in Figure 2.6, bottom panel.

\subsection{Abundance Gradients in External Galaxies}

Abundance gradients expressed in
dex/Kpc are found to be steeper in smaller disks but the
correlation disappears if they are expressed in dex/$R_d$, which means that there is a universal slope per unit scale length (ref).
The gradients are  generally flatter in galaxies with central bars (ref).
The SFR 
is measured mainly from $H_{\alpha}$ emission (Kennicutt, 1998)
and show a correlation with the total surface gas density (HI+$H_{2}$),
in particular the suggested law is that of eq. (1.5).

In the observed gas distributions
differences between field and cluster spirals are found in the sense that cluster spirals have less gas, probably as a consequence of stronger interactions with the environment.Integrated colors of spiral galaxies (Josey \&
Arimoto 1992; Jimenez et al. 1998; Prantzos \& Boissier 2000) indicate 
inside-out formation, as also found for the milky Way.

As an example of abundance gradients in a spiral galaxy we show in Figure 2.7
the observed and predicted gas distribution and abundance gradients for the disk of M101. In this case the gas distribution and the abundance gradients are reproduced with systematically smaller timescales for the disk formation relative to the 
MW (M101 formed faster), and the difference between the timescales of formation of the internal and external regions is 
smaller ($\tau_{M101}=0.75R - 0.5$ Gyr, Chiappini et al. 2003a)

\begin{figure}

\includegraphics[width=4.5in,height=3.0in]{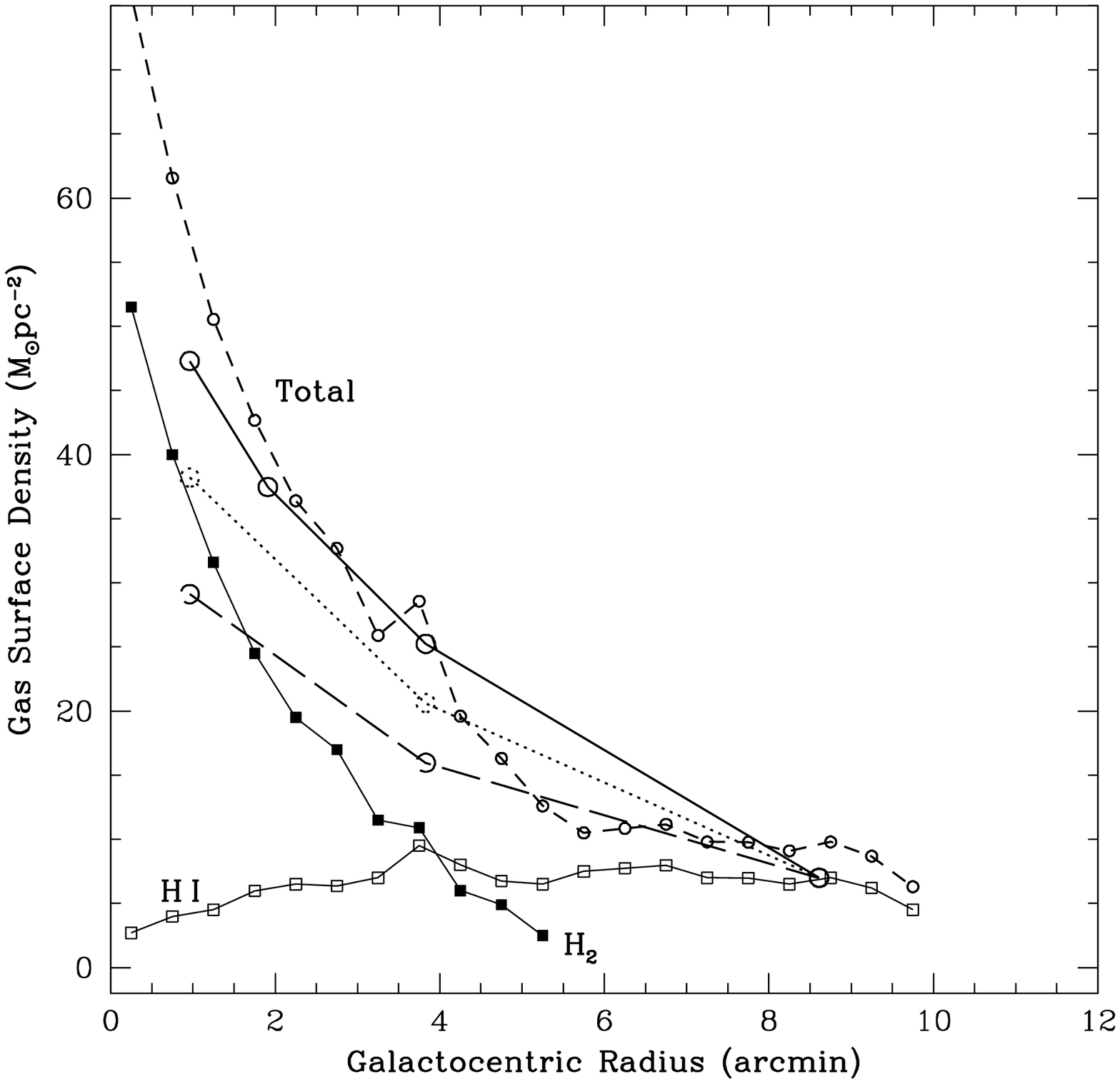}
\includegraphics[width=4.5in,height=3.0in]{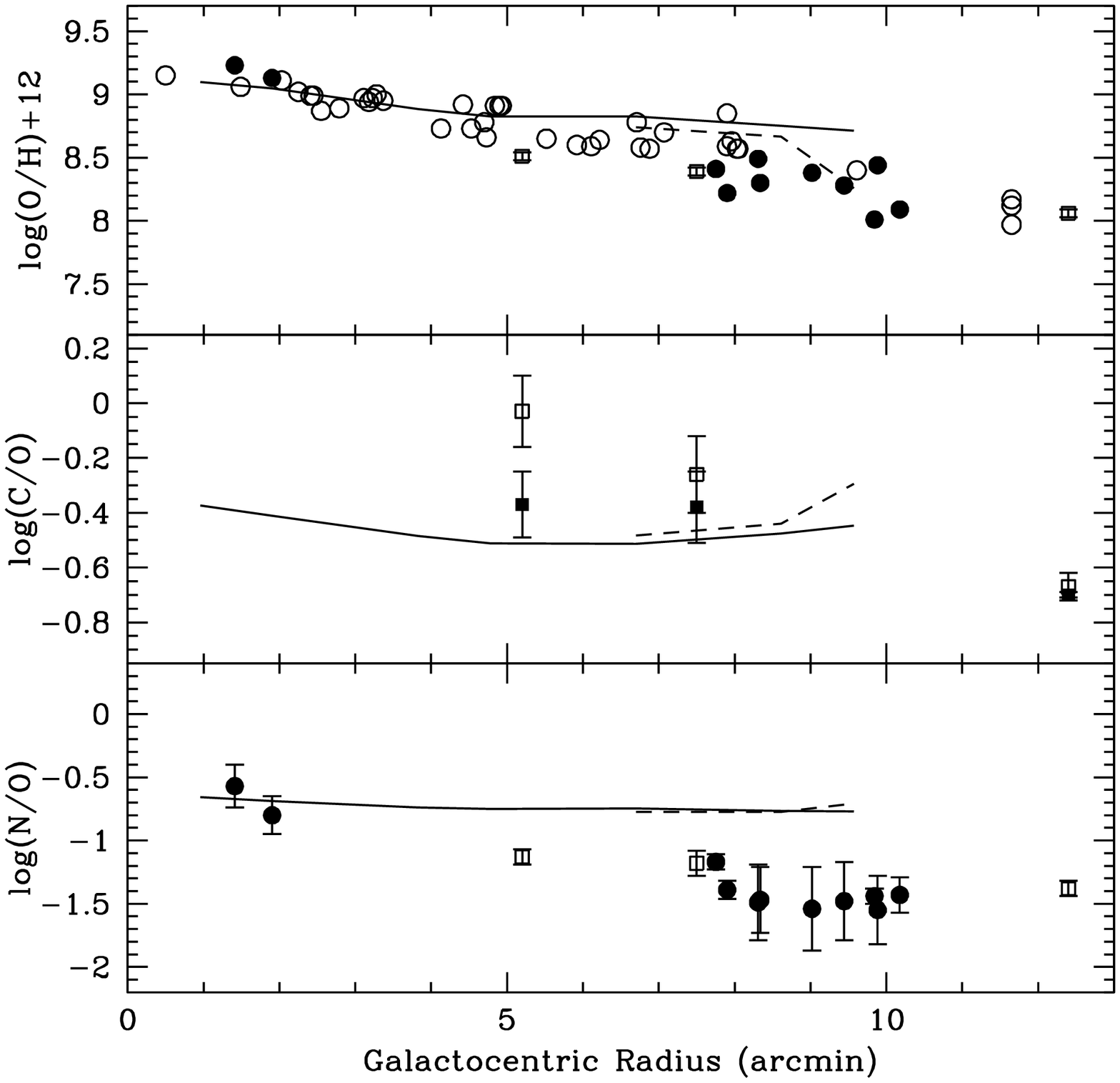}
\hfill
\caption{Upper panel: predicted and observed gas distribution along the disk of M101. The observed HI, $H_2$ and total 
gas are indicated in the Figure.
The large open circles indicate the models: in particular, the open circles connected by a continuous line refer to a model with central surface mass density of 1000$M_{\odot}pc^{-2}$, while the dotted line refers to a model with 800$M_{\odot}pc^{-2}$ and the dashed to a model with  600$M_{\odot}pc^{-2}$.
Lower panel: predicted and observed abundance gradients of C,N,O elements along the disk of M101.The models are the lines and differ for a different threshold density for SF, being larger in the dashed model. All the models are by Chiappini et al. (2003a).}\label{fig} 
\end{figure}

To conclude this section we like to recall a paper by Boissier et al. (2001) 
where a detailed study of the properties of disks is presented. They conclude that more massive disks are redder, more metal rich and more gas-poor  than smaller ones. On the other hand their estimated SF efficiency (defined as the SFR per unit mass of gas) seem to be similar among different spirals: this leads them to conclude that more massive disks are older than less massive ones.

\subsection{How to model the Hubble Sequence}
The Hubble Sequence can be simply thought as a sequence of objects where the SFR proceeds faster in the early than in the late types (see also Sandage, 1986).

We take the Milky Way galaxy, whose properties are best known, as a reference galaxy and we change the SFR relatively to 
the Galactic one, for which we adopt eq. (1.6). The quantity $\nu$ in eq. (1.6) is the efficiency of SF which we assume to be  characteristic of each Hubble type. In the two-infall model for the Milky Way we adopt $\nu_{halo}=2.0 Gyr^{-1}$ and 
$\nu_{disk}=1.0 Gyr^{-1}$ (see Figure 2.1). The choice of adopting a dependence on the total surface mass density for the 
Galactic disk is due to the fact that it helps in producing a SFR strongly varying with the galactocentric distance, as 
required by 
the observed SFR and gas density distribution as well as by the abundance gradients. In fact, the inside-out scenario 
influences the rate at which the gas mass is accumulated by infall at each galactocentric distance and this in turn 
influences the SFR.

For bulges and ellipticals we assume that the SF proceeds like in a burst with very high star formation efficiency, namely:

\begin{equation}
SFR=\nu \sigma^{k}
\end{equation}

with $k=1.0$ for the sake of simplicity; $\nu=10-20 Gyr^{-1}$ (see Matteucci, 1994; Pipino \& Matteucci 2004).

For irregular galaxies, on the other hand, we assume that the SFR proceeds more slowly and less efficiently that in the Milky Way disk, in particular we assume the same SF law as for spheroids but with $0.01 \le \nu (Gyr^{-1}) \le 0.1 $. 
Among irregular galaxies, a special position is taken by the Blue Compact Galaxies (BCG) namely galaxies which have blue colors as a consequence of the fact that they are forming stars at the present time, have small masses, large amounts of gas 
and low metallicities. For these galaxies,  we assume that they suffered on average 
from 1 to 7 short bursts, with  the SF efficiency mentioned above (see Bradamante et al. 1998 and next Lecture).

Finally, dwarf spheroidals are also a special cathegory, characterized by old stars, no gas and low metallicities. For these galaxies we assume that they suffered one long starburst  lasting 7-8 Gyr or at maximum  a couple of extended SF periods, in agreement with their measued Color-Magnitude diagram.
It is worth noting that both ellipticals and dwarf spheroidals should loose most of their gas and therefore one may conclude that galactic winds should play an
important role in their evolution, although ram pressure stripping cannot be 
excluded as a mechanism for gas removal.
Also for these galaxies we assume the previous SF law with $k=1$
and $\nu=0.01-1.0 Gyr^{-1}$. Lanfranchi \& Matteucci, (2003, 2004) developed more detailed models for dwarf spheroidals by adopting the SF history suggested  by the Color-Magnitude diagrams of single galaxies and with the same efficiency of SF as above.
In Figure 2.8 we show the adopted SFRs in different galaxies and in Figure 2.9 the corresponding predicted Type Ia SN rates. For the irregular galaxy, the predicted Type Ia SN rate refers to a specific galaxy, LMC, with a SFR taken from 
observations (see Calura et al. 2003) with an early ans a late burts of SF and low SF in between.

\begin{figure}
\includegraphics[width=4.5in,height=2.5in]{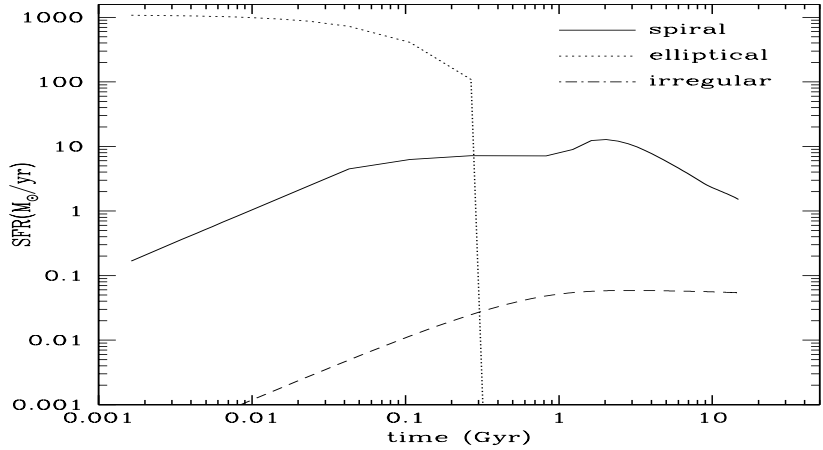}
\hfill
\caption{Predicted SFRs in galaxies of different morphological type. Figure from Calura (2004). Note that for the elliptical galaxy the SF stops abruptly as a consequence of the galactic wind.}\label{fig} 
\end{figure}

\begin{figure}
\includegraphics[width=4.5in,height=2.5in]{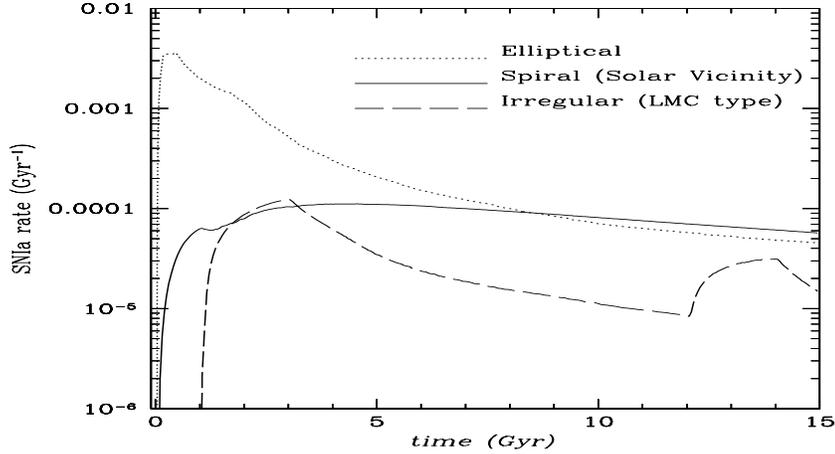}
\hfill
\caption{Predicted Type Ia SN rates for the SFRs of Figure 2.8. Figure from Calura (2004). Note that for the irregular galaxy here the predictions are for the LMC, where a recent SF burst is assumed.}\label{fig} 
\end{figure}

\subsection{Type Ia SN rates in different galaxies}
Following Matteucci \& Recchi (2001) we define the typical timescale for Type Ia SN enrichment  
as {\it the time when the SN rate reaches the maximum}.
In the following we will always adopt the SDS for the progenitors of Type Ia SNe.
A point that is not often understood is that this timescale depends upon the progenitor lifetimes, IMF and SFR and 
therefore is not universal. Sometimes in the literature the typical Type Ia SN timescale is quoted as being universal 
and equal to 1 Gyr, whereas this is just the timescale at which the Type Ia SNe start to be important in the process of 
Fe enrichment in the solar vicinity.

Matteucci \& Recchi (2001) showed that for an elliptical galaxy or a bulge of spiral with a high SFR the timescale for 
Type Ia SN enrichment is quite short, in particular 
$t_{SNIa}= 0.3-0.5$ Gyr.
For a spiral like the Milky Way, in the two-infall model, a first peak is reached at 1.0-1.5 Gyr (the time at which SNeIa become important
as Fe producers (Matteucci and Greggio 1986) while  a second less important 
peak occurs at
$t_{SNIa}=4-5$ Gyr. For an irregular galaxy with a continuous but very low SFR 
the timescale is $t_{SNIa} > 5$ Gyr.

\subsection{Time-delay model for different galaxies}
As we have already seen, the time-delay between the production of oxygen by Type II SNe and that of Fe by Type Ia SNe allows us to explain the 
[X/Fe] vs. [Fe/H] relations
in an elegant way. However, the [X/Fe] vs. [Fe/H] plots depend not only on nucleosynthesis and IMF but also on other model assumptions, such as the SFR, through the absolute Fe abundance ([Fe/H]). Therefore, we should expect a different behaviour in galaxies with different SF histories. In Figure 2.10 we show the predictions of the time-delay model for a spheroid like the Bulge, for the solar vicinity and for a typical irregular magellanic galaxy.

\begin{figure}
\includegraphics[width=4.5in,height=4.0in]{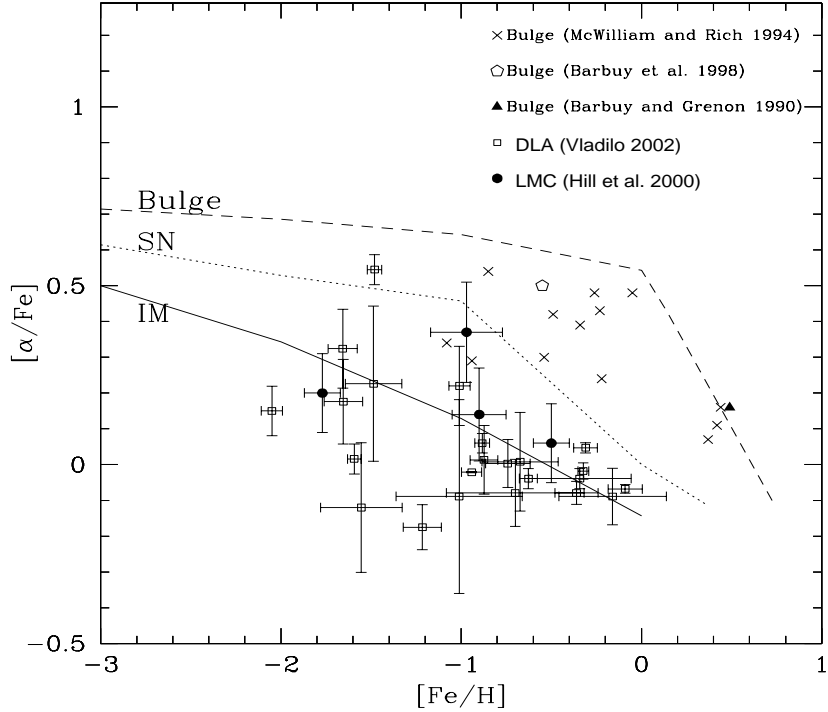}
\hfill
\caption{Predicted [$\alpha$/Fe] ratios in galaxies with different SF histories. The top line represents the predictions for the Bulge or for an elliptical galaxy of the same mass ($\sim 10^{10} M_{\odot}$), the median line represents the prediction for the solar vicinity and the lower line the prediction for an irregular magellanic galaxy. The differences among the various models are in the efficiency of star formation, being quite high for spheroids ($\nu =20Gyr^{-1}$), moderate for the Milky Way ($\nu=1-2 Gyr^{-1}$) and low for irregular galaxies ($\nu = 0.1 Gyr^{-1}$). The nucleosynthesis prescriptions are the same in all objects. The time-delay between the production of $\alpha$-elements and Fe, coupled with the different SF histories produces the differences in the plots. Data for Damped-Lyman-$\alpha$ systems, LMC and Bulge are shown for comparison.}\label{fig} 
\end{figure}

As one can see in this Figure, we predict a long plateau, well above the solar value, for the [$\alpha$/Fe] ratios in the Bulge (and ellipticals), owing to the fast Fe enrichment reached in these systems by means of Type II SNe: when the Type Ia SNe start enriching substantially the ISM, at 0.3-0.5 Gyr, the gas Fe abundance is already solar.
The opposite occurs in Irregulars where the Fe enrichment proceeds very slowly so that when Type Ia SNe start restoring the Fe in a substantial way ($> 3$ Gyr) the Fe in the gas is still well below solar. Therefore, here we observe a steeper slope for the [$\alpha$/Fe] ratio. In other words, we have below solar [$\alpha$/Fe] ratios at below solar [Fe/H] ratios.
This diagram is very important since it allows us to recognize a galaxy type only by means of its abundances, and therefore it can be used to understand the nature of high redshift objects.

\section{Lecture III: interpretation of abundances in dwarf irregulars}

They are rather simple objects with low metallicity
and large gas content, suggesting that they are either young or have
undergone discontinuous star formation activity (bursts) or a continuous but not efficient star formation. 
They are very interesting objects for studying galaxy evolution. In fact, in "bottom-up" 
cosmological scenarios they should be the
first self- gravitating systems to form and they
could also be important contributors to the population 
of systems giving rise
to QSO-absorption lines at high redshift (see Matteucci et al. 1997 
and Calura et al. 2002).

\subsection{Properties  of Dwarf Irregular Galaxies}

Among local star forming galaxies, sometimes referred to as HII galaxies, most are dwarfs.
Dwarf irregular galaxies can be divided into two categories:  Dwarf Irregular (DIG) and Blue Compact galaxies (BCG). These latter have very blue colors due to
active star formation at the present time. 

Chemical abundances in these galaxies are derived from optical emission lines in HII regions. Both DIG and BCG show a distinctive spread in their chemical properties, altough this spread is decreasing with the new more accurate data,  but also a definite mass-metallicity relation. 

From the point of view of chemical evolution,
Matteucci and Chiosi (1983) first studied the evolution of DIG and BCG by means of analytical chemical evolution models including either outflow or infall and concluded that:
closed-box models cannot account for the Z-log G($G= M_{gas}/M_{tot}$) 
distribution
even if the number of bursts varies from galaxy to galaxy
and suggested possible solutions to explain the observed spread. 
In other words, the data show a range of values of the metallicity for a 
given $G$ ratio, and this means that the effective 
yield is lower than that of the 
Simple Model and vary from galaxy to galaxy.

The possible solutions suggested to lower the effective yield were:
\begin{itemize}
\item  {a.} different IMF's

\item {b.} different amounts of galactic wind 

\item {c.} different amounts of infall

\end{itemize}
In Figure 3.1 we show graphically the solutions a), b) and c). Concerning the solution a), one simply varies the IMF, whereas solutions b) and c) have been already descibed (eqs. 1.21 ans 1.23).

\begin{figure}
\includegraphics[width=4.5in,height=2.5in]{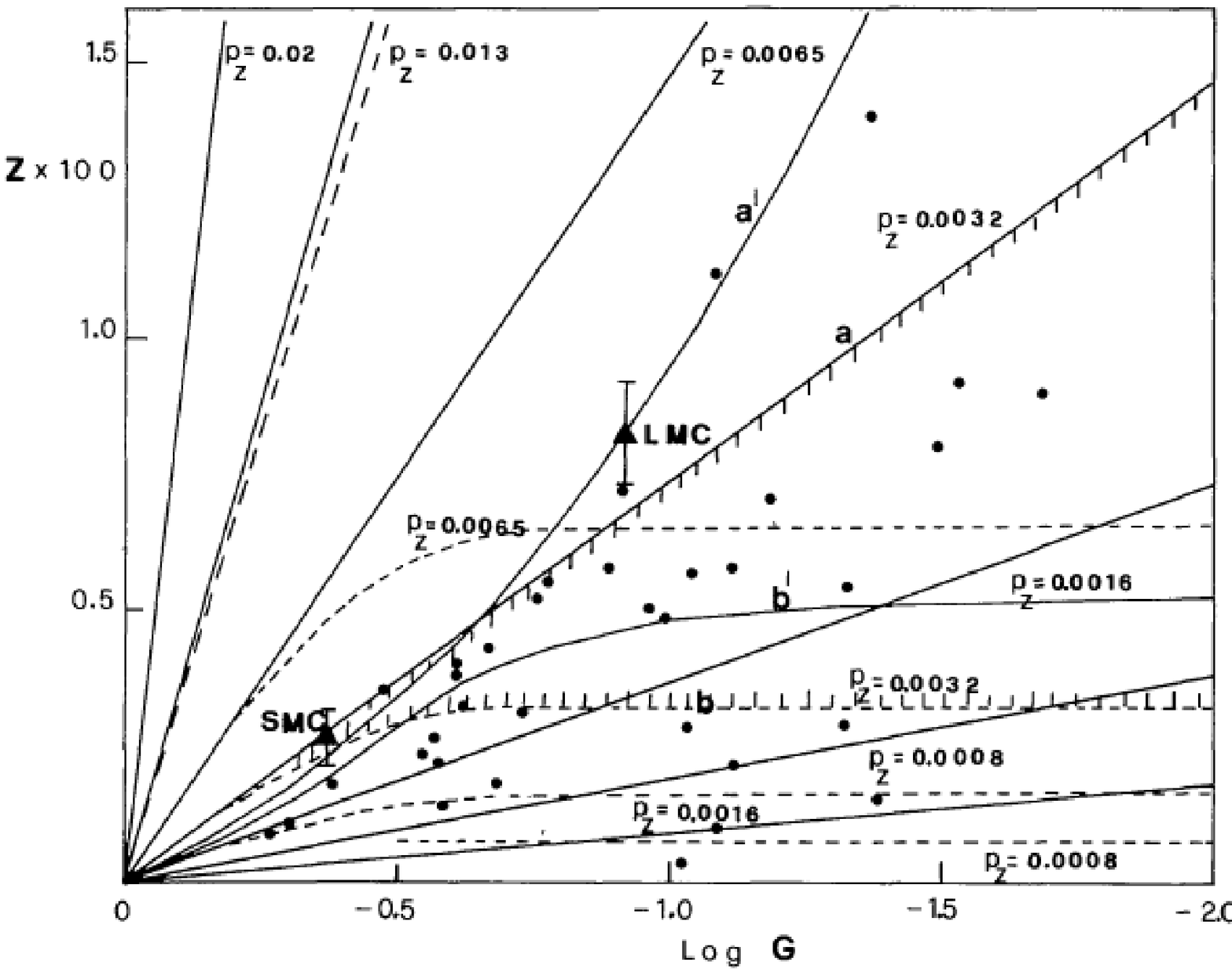}
\includegraphics[width=4.5in,height=2.5in]{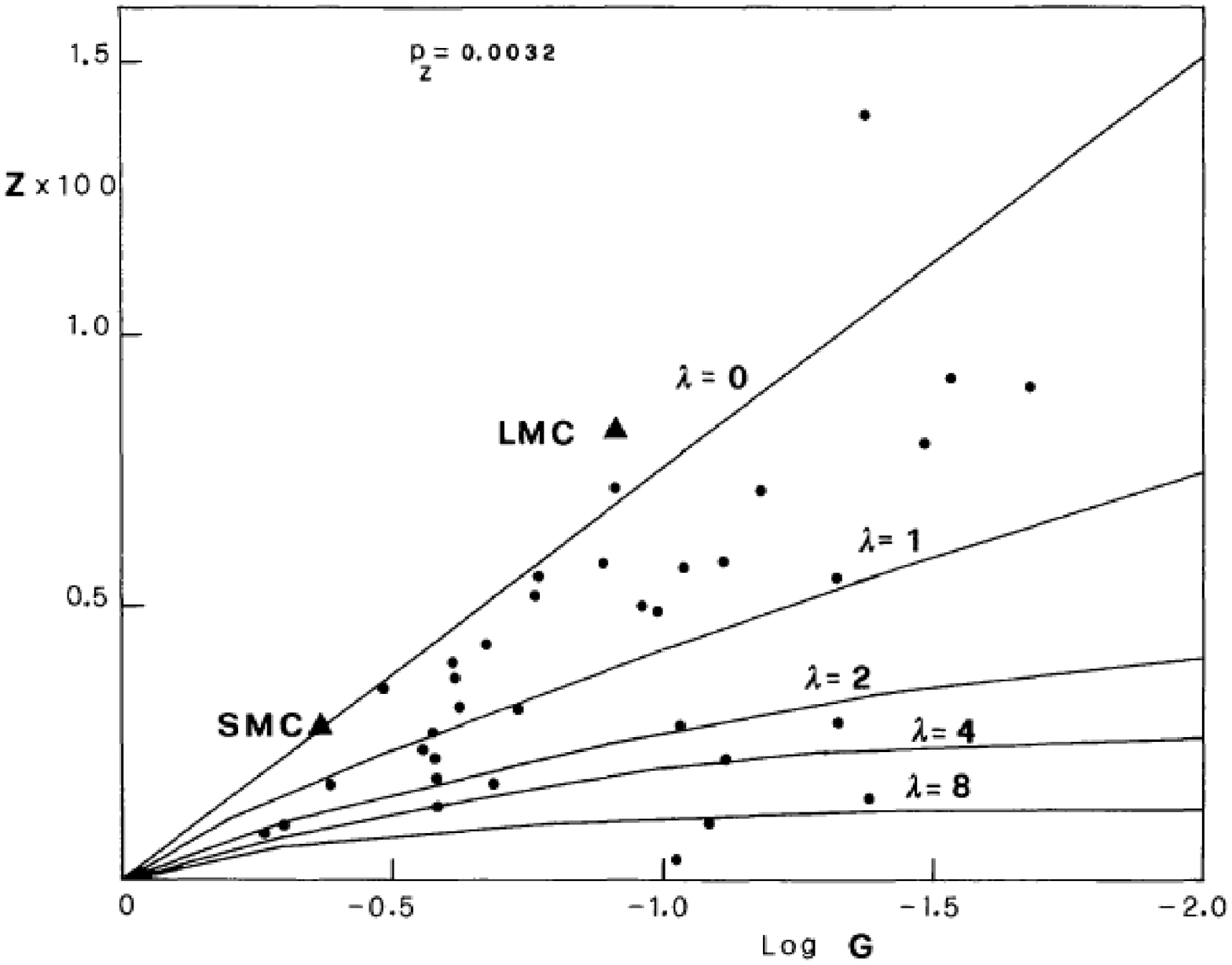}
\includegraphics[width=4.5in,height=2.5in]{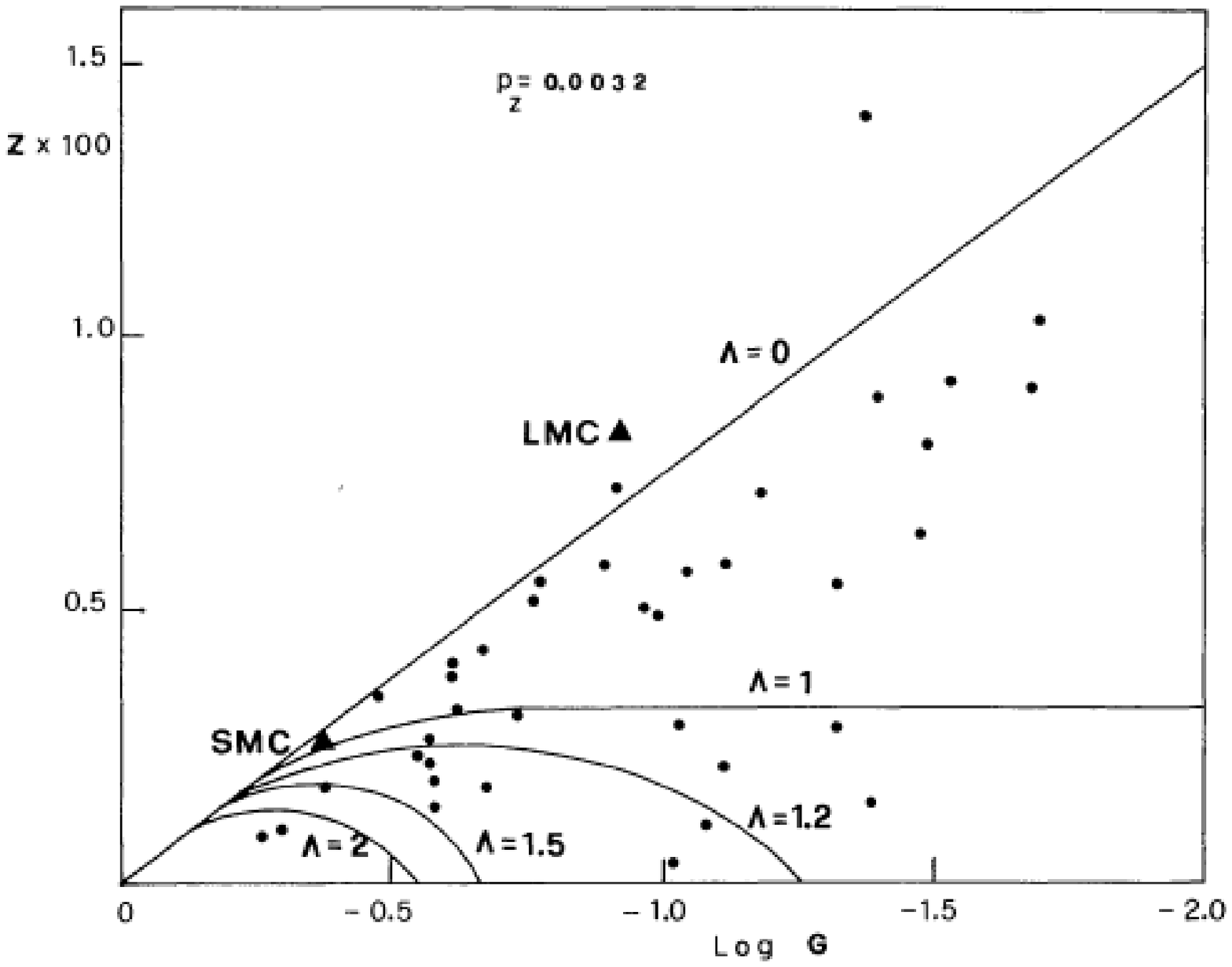}
\hfill
\caption{The Z-$log G$ diagram.Solutions a), b) and c) from top to bottom, to lower the effective yield in DIG and BCG by Matteucci \& Chiosi (1983). Solution a) consists in varying the yield per stellar generation, here indicated by $p_Z$, just by changing the IMF. The solution b) and c) correspond to eqs. (1.21) and (1.23), respectively.}\label{fig} 
\end{figure}

 Later on, Pilyugin (1993) forwarded the idea that the spread observed also in 
other chemical properties properties of these galaxies such as in the He/H vs. O/H  and N/O vs. O/H relations,
can be due to self-pollution of the HII regions, which do not mix efficiently with the surrounding medium,
coupled with ``enriched'' or ``differential'' galactic winds, namely different chemical elements are lost at different rates.
Other models (Marconi et al. 1994; 
Bradamante et al. 1998) followed the suggestions of differential winds and introduced 
the novelty of  
the contribution to the chemical enrichment and energetics of the ISM by 
SNe of different type (II, Ia and Ib).

Another important feature of these galaxies is the mass-metallicity relation.

The existence of a luminosity-metallicity relation in irregulars and BCG
was suggested first by Lequeux et al. (1979), then confirmed by 
Skillman et al. (1989) and extended also to spirals by
Garnett \& Shields (1987).
In particular, Lequeux et al. suggested the relation:

\begin{equation}
M_T = (8.5 \pm 0.4) + (190 \pm 60) Z
\end{equation}
with $Z$ being the global metal content.
Recently, Tremonti et al. (2004) analyzed 53000 local star-forming galaxies in the SDSS (irregulars and spirals). Metallicity was measured from the optical nebular emission lines.
Masses were derived from fitting spectral energy distribution (SED) models. 
The strong optical nebular lines of elements other 
than H are produced by collisionally excited transitions. Metallicity was then determined by fitting simultaneously the most prominent emission lines ([OIII], 
$H_\beta$, [OII], $H_\alpha$, [NII], [SII]).
Tremonti et al. (2004) derived a relation indicating that 12+log(O/H) is
increasing steeply from $M_{*}$ going from $10^{8.5}$ 
to $10^{10.5}$ but flattening for $M_{*} > 10^{10.5}$.

In particular, the Tremonti et al. relation is:
\begin{equation}
12 + log(O/H) = -1.492 + 1.847(log M_{*}) - 0.08026(log M_{*})^{2}.
\end{equation}

This relation extends to higher masses the mass-metallicity relation found for star forming dwarfs and contains very important information on the physics governing galactic evolution.
Even more recently, Erb et al. (2006) found the same mass-metallicity relation 
for star-forming galaxies at redshift z$>$2, with an offset from the local 
relation of $\sim 0.3$ dex. They used $H_{\alpha}$ and [NII] spectra.
In Figure 3.2 we show the figure from Erb et al. (2006) for the mass-metallicity relation at high redshift which includes the relation of Tremonti et al. (2004) for the local mass-metallicity relation.

\begin{figure}
\includegraphics[width=4.5in,height=4.0in]{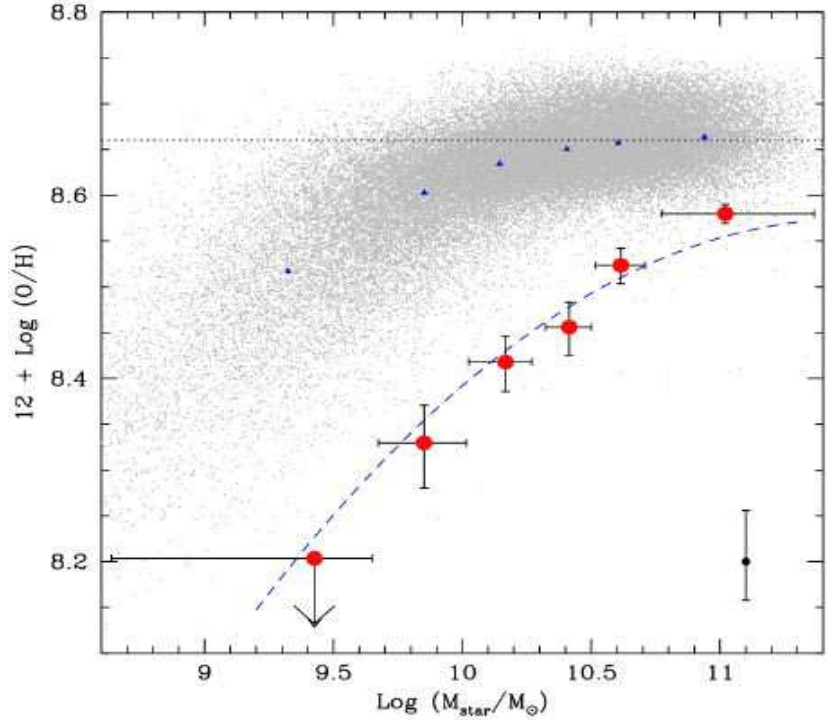}
\hfill
\caption{Figure 3 from Erb et al. (2006) showing the mass-metallicity relation for star forming galaxies at high redshift. The data from Tremonti et al. (2004) are also shown.}\label{fig} 
\end{figure}

The most simple interpretation of the mass-metallicity relation is  that the 
effective yield increases with galactic mass. This can be achieved in several
ways, as shown in Fig. 3.1.: either by 
changing the IMF or the stellar yields as a function of galactic mass, or by assuming that the galactic wind is less efficient in more massive systems, or that the infall rate is less efficient in more massive systems.
One of the  most common interpretations of the mass-metallicity relation is that the effective yield changes because 
of the occurrence of galactic winds, which should be more important in small systems.
Evidences for galactic winds exist for dwarf irregular galaxies, as we will see next. 

\subsection{Galactic Winds}
Papaderos et al.  (1994) estimated a galactic wind flowing 
at a velocity of 1320 Km/sec
for the irregular dwarf VIIZw403. The escape velocity estimated for
this galaxy is $\simeq$ 50 Km/sec.
Lequeux et al. (1995) suggested a galactic wind in Haro2=MKn33 flowing at 
a velocity of $\simeq 200 Km/sec$, also larger that the escape velocity of this object. More recently, Martin (1996;1998) found also supershells in 12 dwarfs, including
IZw18, which imply gas outflow. Martin (1999) concluded that the galactic wind 
rates are several times the SFR.
Finally, the presence of metals in the ICM 
(revealed by X-ray observations) and in the IGM 
(Ellison et al. 2000) represents a clear indication of the fact that galaxies lose their metals. However, we cannot exclude that the gas with metals is lost 
also by ram pressure stripping, especially in galaxy clusters.

In models of chemical evolution of dwarf irregulars (e.g. Bradamante et al. 1998) the feedback effects are taken into account and the condition for the development of a wind is:

\begin{equation}
(E_{th})_{ISM} \ge E_{Bgas}
\end{equation}
namely, that the thermal energy of the gas is larger or equal to its 
binding energy.
The thermal energy of gas due to SN and stellar wind heating is:
\begin{equation}
(E_{th})_{ISM}=E_{th_{SN}}+ E_{th_{w}}
\end{equation}

with the contribution of SNe being:
\begin{equation}
E_{th_{SN}}= \int^{t}_{0}{\epsilon_{SN}R_{SN}(t^{`})dt^{`}},
\end{equation}

while the contribution of stellar winds is:
\begin{equation}
E_{th_{w}}=\int^{t}_{0}\int^{100}_{12}{ \varphi(m) \psi(t^{`}) \epsilon_{w}dm
dt^{`}}
\end{equation}

with
$\epsilon_{SN}= \eta_{SN}\epsilon_{o}$ and $\epsilon_o=10^{51}$erg 
(typical SN energy),
and
$\epsilon_{w}= \eta_{w}E_{w}$ with $E_{w}= 10^{49}$erg 
(typical energy injected by a $20M_{\odot}$ star taken as representative).
$\eta_{w}$ and $\eta_{SN}$ are two free parameters and indicate the 
efficiency of
energy transfer from stellar winds and SNe into the ISM, respectively,  
quantities still largely unknown.
The total mass of the galaxy is expressed as 
$M_{tot}(t)=M_{*}(t)+M_{gas}(t)+M_{dark}(t)$
with $M_L(t)=M_{*}(t)+M_{gas}(t)$
and the binding energy of gas is:

\begin{equation}
E_{Bgas}(t)=W_L(t)+W_{LD}(t)
\end{equation}

with:
\begin{equation}
W_L(t)=-0.5G{ M_{gas}(t) M_L(t) \over r_L}
\end{equation}
which is the potential well due to the luminous matter and
with:
\begin{equation}
W_{LD}(t)= -Gw_{LD}{M_{gas}(t) M_{dark} \over r_L}
\end{equation}
which represents the potential well due to the interaction between dark and 
luminous matter,
where $w_{LD} \sim {1 \over 2\pi} S(1+1.37S)$,
with $S= r_L/r_{D}$, being the ratio between the galaxy effective radius and the radius of the dark matter core. 
The typical model for a BCG has a luminous mass of $10^{8}-10^{9}M_{\odot}$,  a dark matter halo ten times larger than the luminous mass and various values for the parameter $S$. The galactic wind in these galaxies develops easily but it carries out mainly metals so that the total mass lost in the wind is small.

\subsection{Results on DIG and BCG from purely chemical models}

Purely chemical models (Bradamante et al. 1998, 
Marconi et al. 1994) for DIG and BCG have been computed in the last years
by varying  the number of bursts, the time of occurrence of bursts 
$t_{burst}$, the star 
formation efficiency, the type of galactic wind (differential or normal), the 
IMF and the 
nucleosynthesis prescriptions.
The best model  of Bradamante et al. (1998) suggests that
the number of bursts should be 
$N_{bursts} \le 10$, the SF efficiency should vary from 0.1 to 
0.7 $Gyr^{-1}$ for either Salpeter or Scalo (1986) IMF (Salpeter IMF 
is favored).  Metal enriched winds are favored. 
The results of these models also suggest that
SNe of Type II dominate the chemical evolution and energetics of these 
galaxies, whereas stellar winds are negligible.
The predicted [O/Fe] ratios tend to
be overabundant relative to the solar ratios, owing 
to the predominance of Type II SNe during the bursts, in agreement with 
observational data (see Figure 3.5 upper panel). Models with strong 
differential winds and  $N_{burst}$=10 - 15 can however give rise to 
negative [O/Fe] ratios. The main difference between DIGs and BCGs, in these 
models, is that the BCGs suffer a present time burst, whereas the DIGs are 
in a quiescent phase.

In Figure 3.3 we show some of the results of Bradamante et al. (1998) 
compared with data on BCGs: it is evident from the Figure that the spread in 
the chemical properties can be simply reproduced by different SF efficiencies, 
which translate into different wind efficiencies.

\begin{figure}
\includegraphics[width=4.5in,height=2.5in]{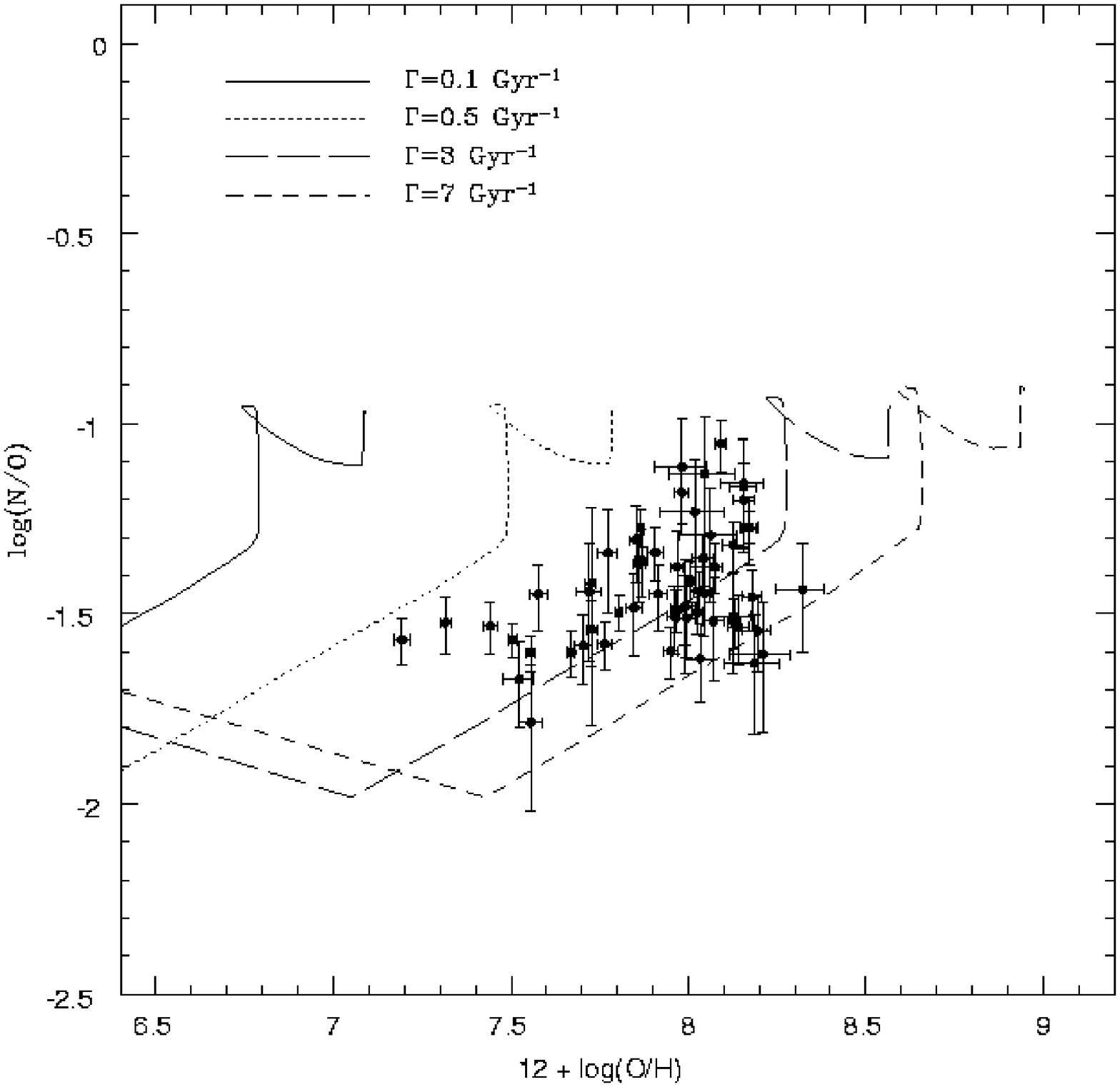}
\includegraphics[width=4.5in,height=2.5in]{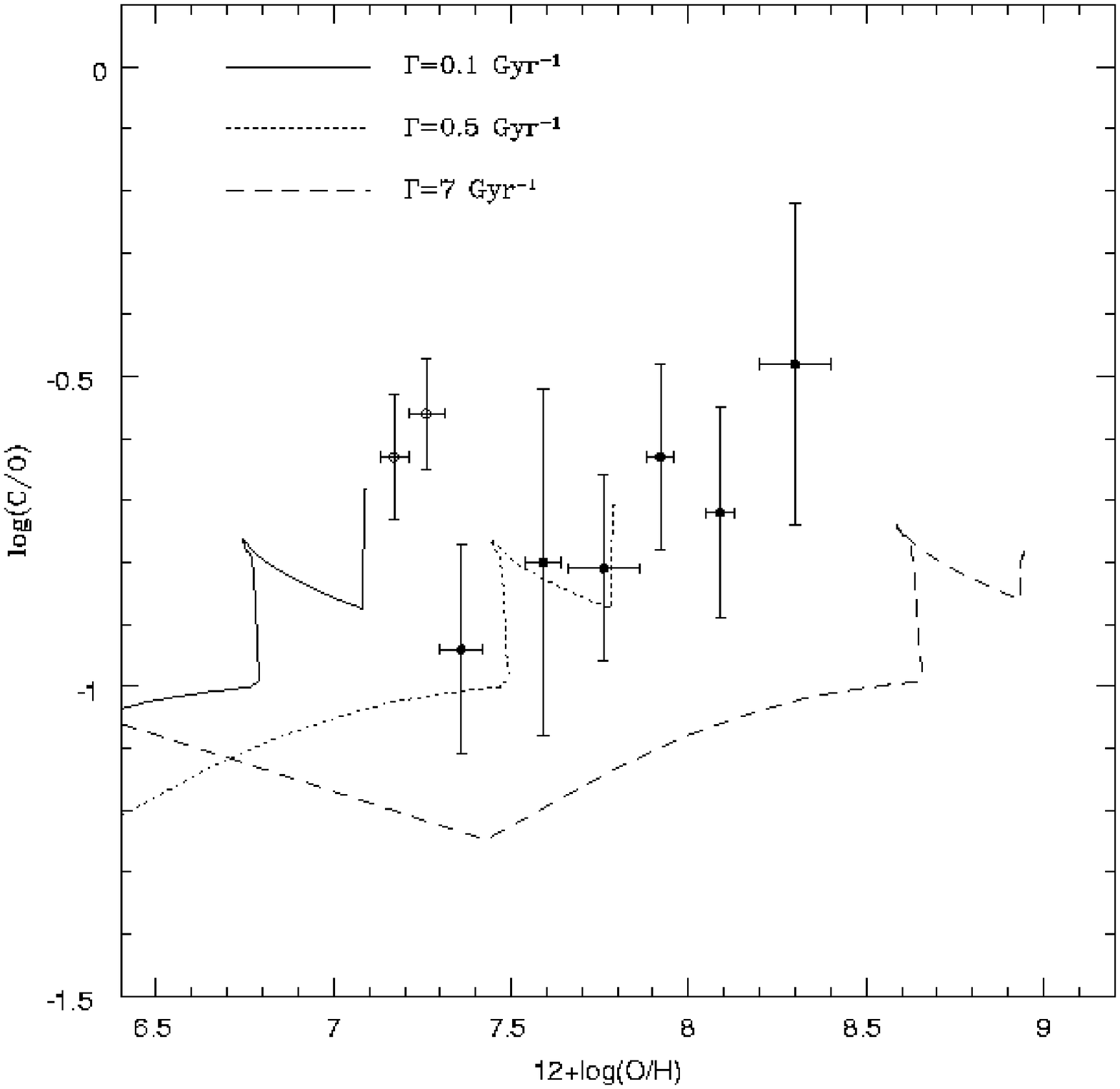}
\hfill
\caption{Upper panel : predicted Log(N/O) vs. 12 + log(O/H) for a model with 3 bursts of SF separated by quiescent periods and different SF efficiencies here indicated with $\gamma= \nu$. Lower panel: predicted log(C/O) vs. 12 + log(O/H). The data in both panels are from Kobulnicky and Skillman (1996). The models assume a dark matter halo ten times larger than the luminous mass and S=0.3 ( Bradamante et al. 1998, see text).}\label{fig} 
\end{figure}

In Fig 3.4 we show the results of the chemical evolution models of
Henry et al. (2000). These models take into account exponential infall but not outflow. They suggested that the SF efficiency  in extragalactic HII regions must have been low and that this effect coupled with the primary N production from intermediate mass stars can explain the plateau in log(N/O) observed at low 12+log(O/H). Henry et al. (2000) also concluded that $^{12}C$ is mainly produced in massive stars (yields by Maeder 1992) whereas $^{14}N$ is mainly produced in intermediate mass stars (yields by HG97). This conclusion, however, should be tested also on the abundances of stars in the Milky Way, where the flat behaviour of [C/Fe] vs. [Fe/H] from [Fe/H] =-2.2 up to [Fe/H]=0 suggest a similar origin for the two elements, namely partly from massive stars and mainly from low and intermediate mass ones (Chiappini et al.  2003b).

\begin{figure}
\includegraphics[width=4.5in,height=3.0in]{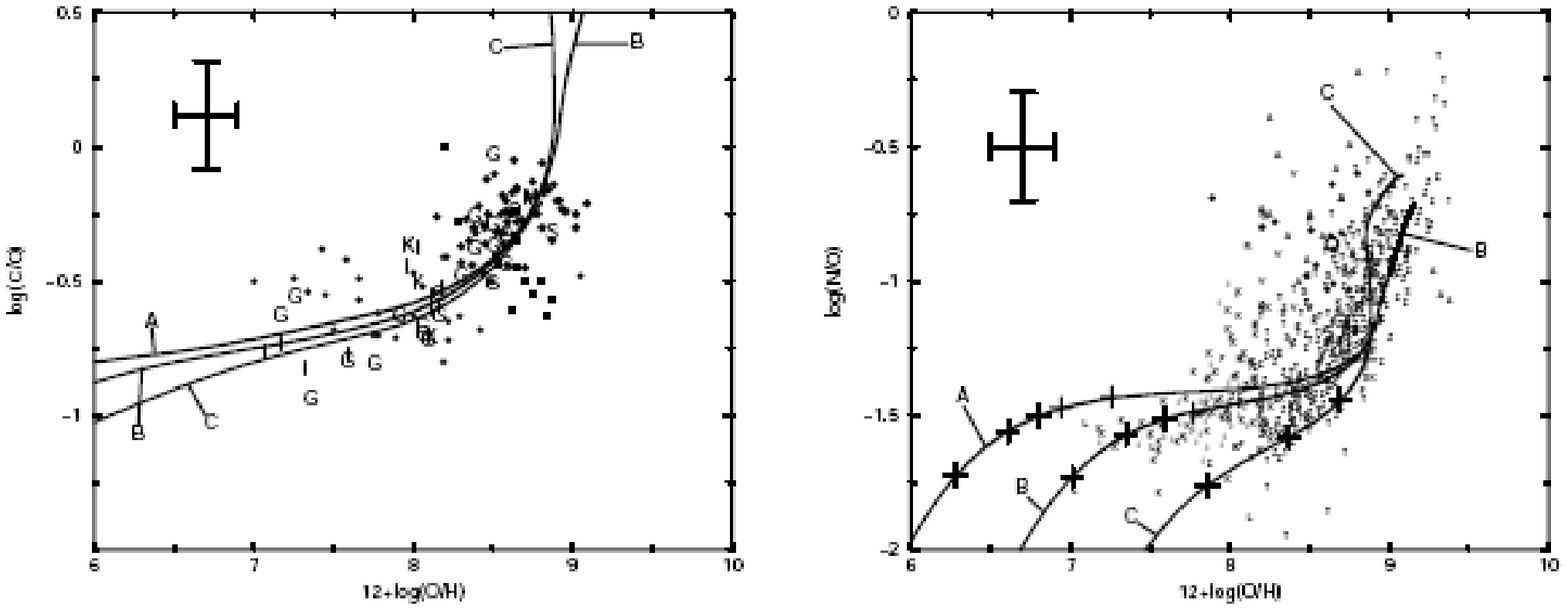}
\hfill
\caption{Figure from Henry et al. (2000): a comparison between numerical models and data  for extragalactic HII regions and stars (filled circles, filled boxes and filled diamonds); M and S mark the position of the Galactic HII regions and the Sun, respectively. Their best model is model B with an efficiency of SF of  $\nu=0.03$.}\label{fig} 
\end{figure}

Concerning the [O/Fe] ratios we show results from Thuan et al. (1995) in Figure 
3.5, where it is evident that generally BCGs have overabundant [O/Fe] ratios.

\begin{figure}
\includegraphics[width=5in,height=3.0in]{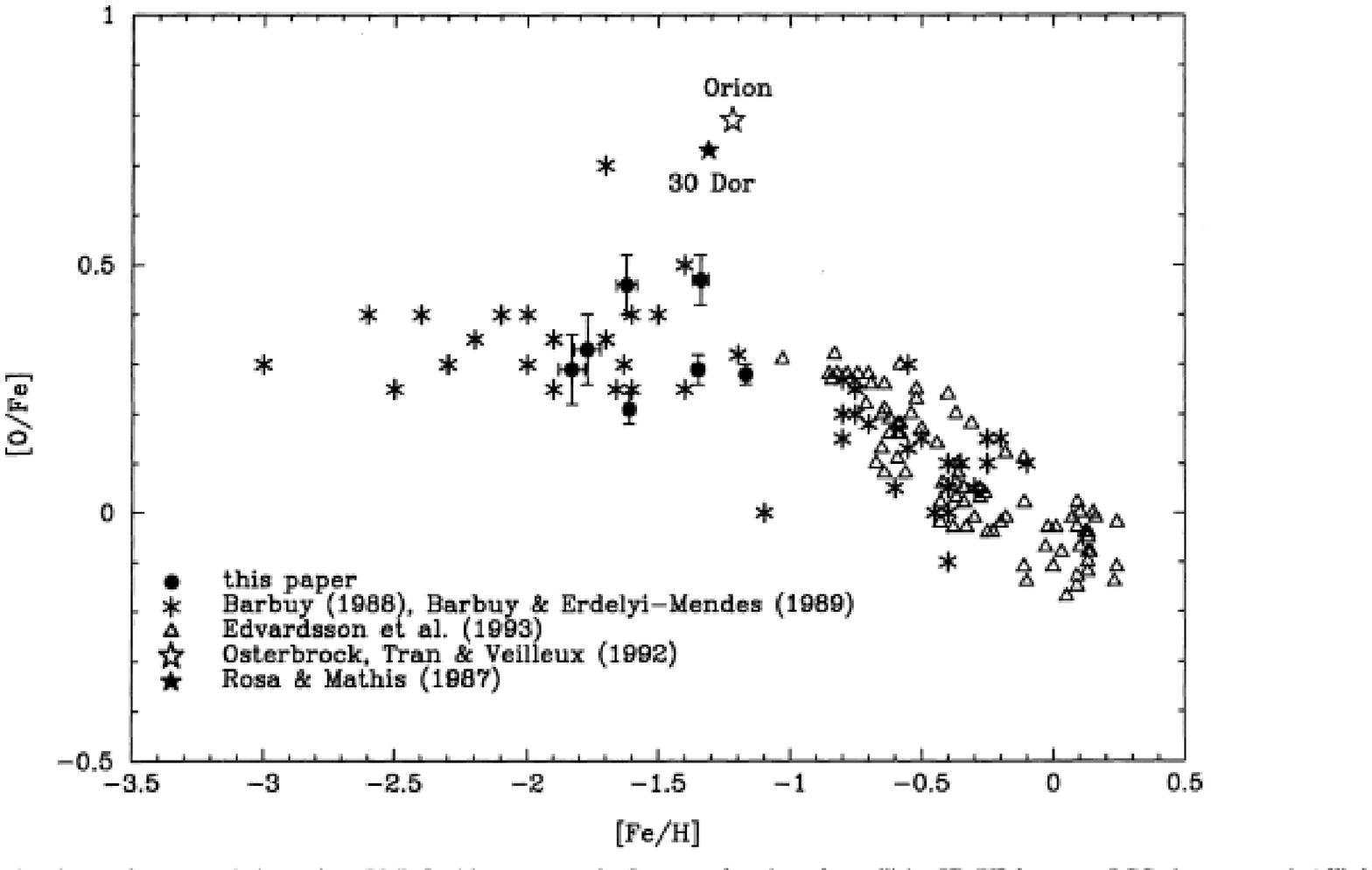}
\includegraphics[width=5in,height=3.5in]{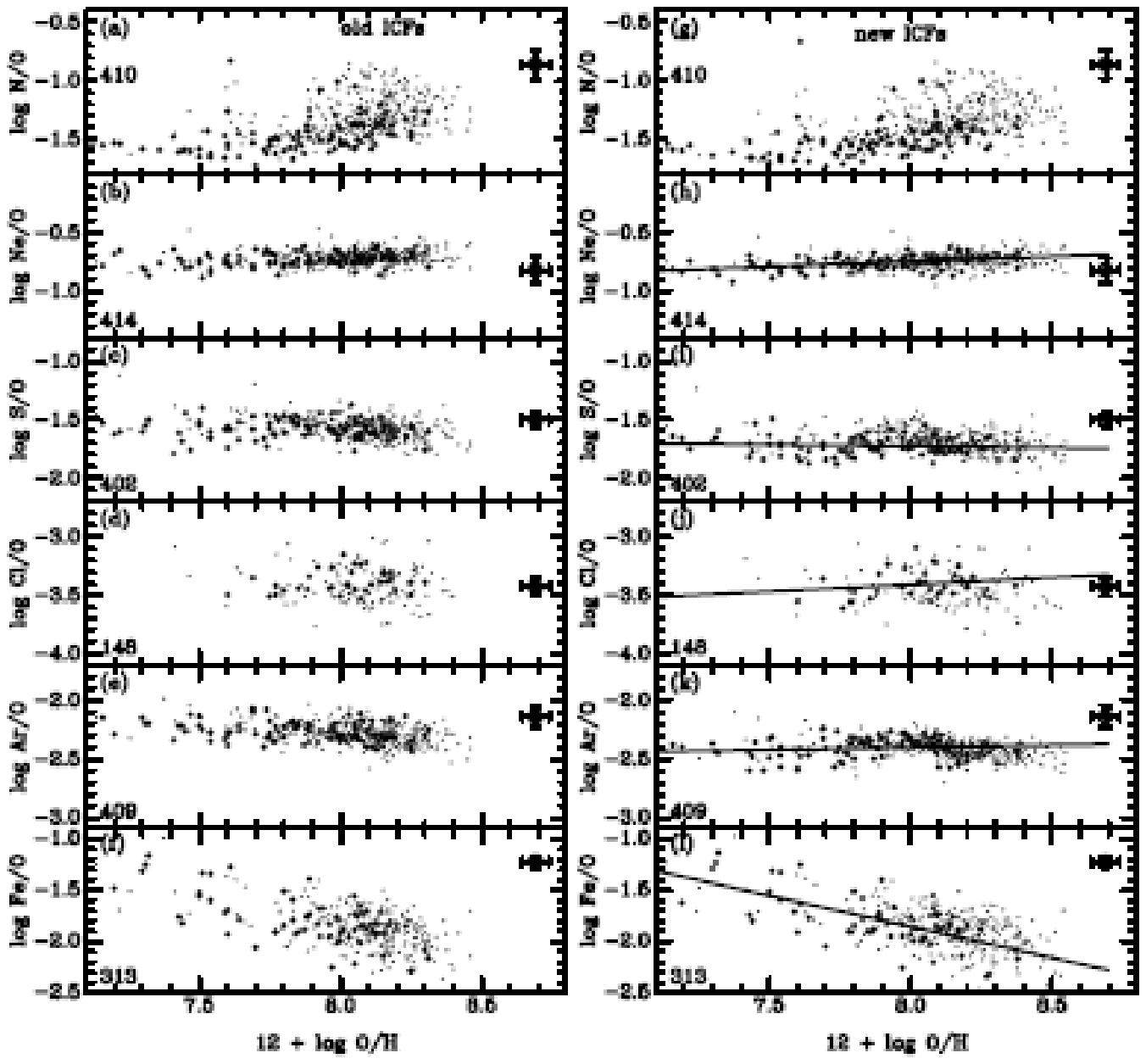}
\hfill
\caption{Upper panel: [O/Fe] vs. [Fe/H] observed in a sample of BCGs by Thuan et al. (1995) (filled circles), open triangles and asterisks are disk and halo stars shown for comparison.Figure from Thuan et al. (1995). Lower panel: new data from Izotov et al. (2006). The large filled circles represent the BCGs whereas the dots are the SDSS galaxies. Abundances in the left panel are calculated as in Thuan et al. (1995) whereas those in the right panel are calculated as in Izotov et al. (2006) (see original papers for details). Figure from Izotov et al. (2006).}\label{fig} 
\end{figure}

Very recently, an extensive study from SDSS of chemical abundances
from emission lines in a sample of 310 metal poor emission line galaxies 
appeared (Izotov et al. 2006).
The global metallicity in these galaxies ranges from 
$\sim 7.1 (Z_{\odot}/30)$ to 
$\sim8.5(0.7Z_{\odot})$. The SDSS sample is merged with 109 BCGs 
containing extremely low metallicity objects.
These data, shown in Figure 3.5 lower panel,  substantially confirm previous ones, showing how $\alpha$-elements do not depend on the O abundance suggesting a common origin for these elements in stars with $M > 10 M_{\odot}$, except for a slight increase of Ne/O with metallicity which is interpreted as due to a moderate dust depletion of O in metal rich galaxies. An important finding is that all the studied galaxies are found to have
log$(N/O) > -1.6$, which indicates that none of these galaxies is a truly young object, unlike the DLA systems at high redshift which show
a log$(N/O) \sim -2.3$.

\subsection{Results from Chemo-Dynamical models: IZw18}

IZw18 is the most metal poor local galaxy, thus  resembling to 
a primordial object. Probably it did not experience more than two bursts of 
star formation including the present one. The age of the oldest stars in this galaxy is still uncertain, although recently Tosi et al. (2006) suggested an age possibly $>2$ Gyr.
The oxygen abundance in IZW18 is 12+log(O/H)= 7.17-7.26, 
$\sim$ 15-20 times lower than the solar oxygen  (12+ log(O/H)= 8.39, Asplund et al. 2005)
and log N/O= -1.54/ -1.60 (Garnett et al. 1997).

Recently, FUSE provided abundances also for HI in IZw18:
the evidence is that the abundances in the HI are lower than in the HII (Aloisi et al. 2003; Lecavelier des Etangs et al. 2003). In particular, Aloisi et al. (2003) found the largest difference relative to the HII data.

Chemo-dynamical (2-D) models (Recchi et al. 2001) studied first the case of IZw18 with only one burst at the present time and concluded that
the starburst triggers a galactic outflow. In particular, 
the metals leave the galaxy more 
easily than the unprocessed gas and among the enriched material the
SN Ia ejecta leave the galaxy more easily than other ejecta.
In fact, Recchi et al. (2001) had reasonably assumed that Type Ia SNe can transfer almost all of their energy to the gas, since they explode in an already hot and rarified medium after the SN II explosions. As a consequence of this, 
they predicted that the [$\alpha$/Fe] ratios in the gas
inside the galaxy should be larger than the [$\alpha$/Fe] ratios in the gas
outside the galaxy.
At variance with previous studies, they found that most of the metals 
are already in the cold gas phase after 8-10 Myr since the
superbubble does 
not break immediately and thermal conduction can act efficiently.
In the following, Recchi et al. (2004) extended the model to a two-burst case, always with the aim of reproducing the characteristics of IZw18.
The model  
well reproduces the chemical properties of IZw18 with a relatively 
long episode of SF lasting 270 Myr plus a recent burst of SF still going on.
In Figure 3.6 we show the predictions of Recchi et al. (2004) for the 
abundances in the HII regions of IZW18 and in Figure 3.7 those for the
HI region, showing a little difference between the HII and HI abundances, more in agreement with the data of Lecavelier des Etangs
et al. (2004).
\begin{figure}
\includegraphics[width=5in,height=3in]{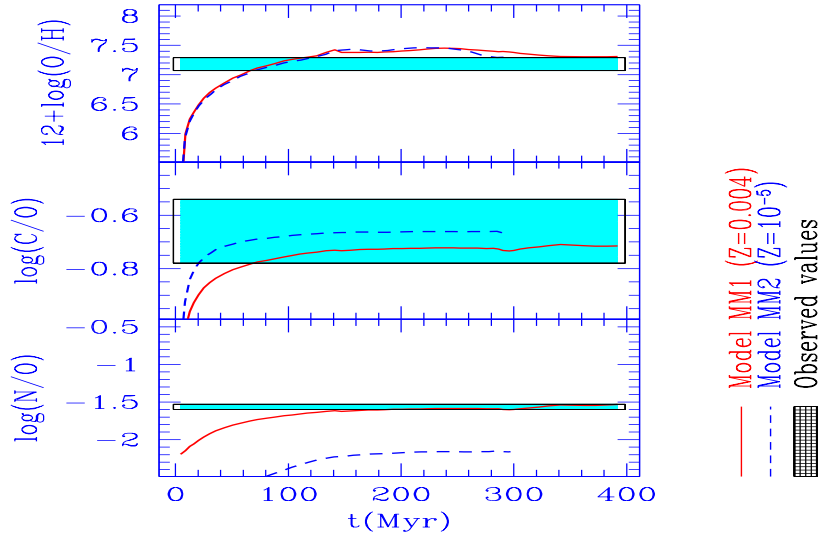}
\hfill
\caption{Figure from  Recchi et al. (2004): predicted abundances for the HII region in IZw18 (dashed lines represent a model  adopting the yields of Meynet \& Maeder (2002) for $Z=10^{-5}$, whereas the continuous line refers to a higher metallicity (Z=0.004).Observational data are represented by the shaded areas.}\label{fig} 
\end{figure}

\begin{figure}
\includegraphics[width=5in,height=3in]{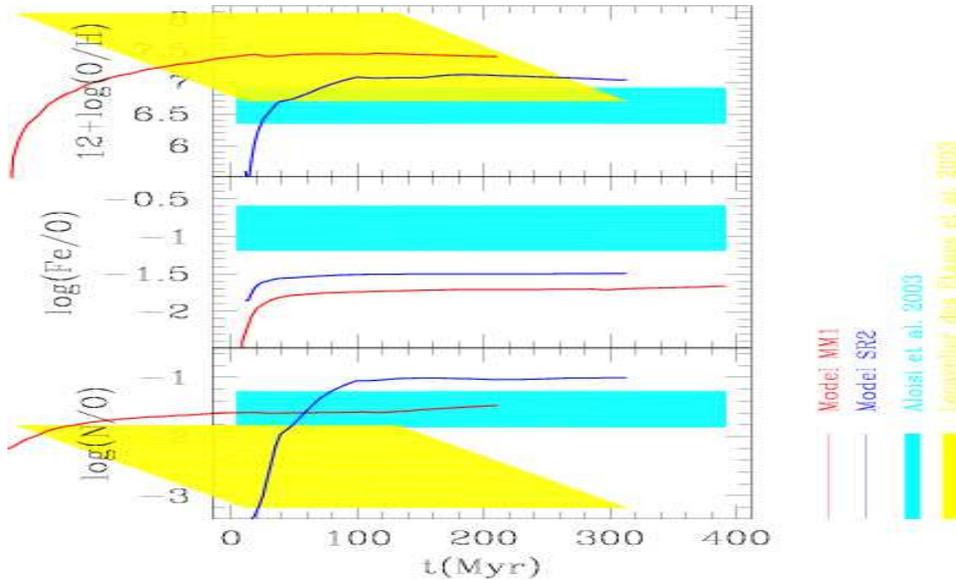}
\hfill
\caption{Figure from  Recchi et al. (2004): predicted abundances for the HI region. The models are the same as in Figure 3.6. Observational data are represented by the shaded areas. The upper shaded area in the panel for oxygen and the lower shaded area in the panel  for N/O represent the data of Lecavelier des Etangs et al. (2003).}\label{fig} 
\end{figure}

\section{Lecture IV: Elliptical galaxies-Quasars- ICM Enrichment}

\subsection{Ellipticals}
We recall here some of the most important properties of ellipticals or early type galaxies (ETG) which are systems made of old stars with no gas and no ongoing SF. The metallicity of ellipticals is measured only by means of metallicity indeces obtained from their integrated spectra which are very similar to those of K giants. In order to pass from metallicity indices to [Fe/H] one needs then to adopt a suitable calibration often based on population synthesis models (Worthey, 1994).
We also summarize the most common scenarios for the formation of ellipticals.

\subsection{Chemical Properties}
The main properties of the stellar populations in ellipticals are:
\begin{itemize}
\item There exist the well-known Color-Magnitude and Color - $\sigma_o$ 
(velocity dispersion) relations indicating that the integrated 
colors become redder 
with increasing luminosity and mass (Faber 1977; Bower et al. 1992). These relations are interpreted as a metallicity effect, although a well known degeneracy exists between metallicity and age of the stellar populations in the integrated colors (Worthey 1994).

\item The index $Mg_{2}$ is normally used as a metallicity indicator since it does not depend much upon the age of stellar populations.
There exists for ellipticals a well defined $Mg_{2}$--$\sigma_o$ relation, 
equivalent to the already discussed mass-metallicity relation for star forming 
galaxies (Bender et al. 1993; Bernardi et al. 1998; Colless et al. 1999).

\item Abundance gradients in the stellar populations inside ellipticals are 
found
(Carollo et al. 1993; Davies et al. 1993).
Kobayashi \& Arimoto (1999) derived the average gradient for ETGs from a large compilation of data and this is: $\Delta [Fe/H] / \Delta r 
\sim -0.3$, with the average metallicity in ETGs of $<[Fe/H]>_{*} \sim -0.3$dex
 (from -0.8 to +0.3 dex).

\item  A very important characteristic of ellipticals is that  their central dominant stellar population (dominant in the visual light) shows an overabundance, relative to the Sun, of the Mg/Fe ratio,
$<[Mg/Fe]>_{*} > 0$ (from 0.05 to + 0.3 dex)
(Peletier 1989;
Worthey et al. 1992; Weiss et al. 1995; Kuntschner et al. 2001).

\item In addition, the overabundance increases with increasing galactic mass and
luminosity,
$<[Mg/Fe]>_{*} vs.\, \sigma_o$,  (Worthey et al. 1992;
Matteucci 1994; Jorgensen 1999; Kuntschner et al. 2001).

\end{itemize}

\subsection{Scenarios for galaxy formation}
The most common ideas on the formation and evolution of ellipticals can be summarized as:

\begin{itemize}

\item  they formed by an early monolithic collapse of a gas cloud or early 
merging of lumps of gas where dissipation plays a fundamental role 
(Larson 1974;
Arimoto \& Yoshii 1987; Matteucci \& Tornamb\`e 1987). In this model SF proceeds very intensively until a galactic wind is developed and SF stops after that. The galactic wind is devoiding the galaxy from all its residual gas.

\item  They formed by means of intense bursts of star formation in merging 
subsystems made of gas
(Tinsley \& Larson 1979). 
In this picture SF stops after the last burst and 
gas is lost via ram pressure stripping or galactic wind.

\item  They formed by early merging of lumps containing gas and stars in which 
some dissipation is present (Bender et al. 1993).

\item  They formed and continue to form in a wide 
redshift range and preferentially at late epochs by merging of early formed 
stellar (e.g. Kauffmann et al. 1993;1996).

\end{itemize}

\begin{figure}
\includegraphics[width=4.5in,height=4.0in]{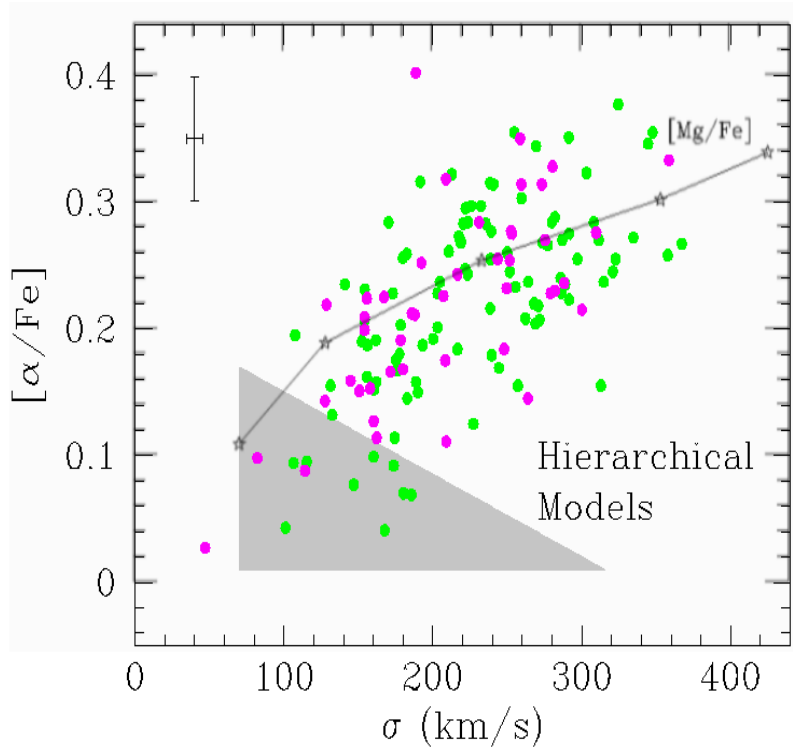}
\hfill
\caption{The relation [$\alpha$/Fe] vs. velocity dispersion (mass)  for ETGs.
Figure adapted from Thomas et al. (2002).The continuous line represents the prediction of the model by Pipino \& Matteucci (2004). The shaded area represents the prediction of hierarchical models for the formation of ellipticals.The symbols are the observational data.}\label{fig} 
\end{figure}
Pipino \& Matteucci (2004), by means of recent revised monolithic models
taking into account the development of a galactic wind (see Lecture III),
computed the relation [Mg/Fe] versus mass (velocity  dispersion) and compared 
it with the data by Thomas et al. (2002). Thomas (1990) already showed how 
hierarchical semi-analitycal models cannot reproduce the observed [Mg/Fe] vs. 
mass trend, since in this scenario massive ellipticals have longer periods of 
star formation than smaller ones. In Figure 4.1, the original figure from 
Thomas et al. (2002) is shown, where  we have plotted also our predictions.
In the Pipino \& Matteucci (2004) model it is assumed that the most massive 
galaxies assemble faster and form stars faster than less massive ones.  The adopted IMF is the Salpeter one. In other words, more massive ellipticals seem to be older than less massive ones, in agreement with what found for spirals (Boissier et al. 2001). 
In particular, in order to explain the observed $<[Mg/Fe]>_{*} > 0$
in giant ellipticals the dominant stellar population should have formed 
on a time scale no longer than 3-5 $\cdot 10^{8}$ yr (Weiss et al. 1995; Pipino \& Matteucci 2004).

\subsection{Ellipticals-Quasars connection}

We know now that most if not all massive ETGs are hosting an AGN for sometime during their life. Therefore, there is a strict link between the quasar activity and the evolution of ellipticals.
\subsection{The chemical evolution of QSOs}
It is very interesting to study the chemical evolution of QSOs by means of the broad emission lines in the QSO region.
The first studies by Wills et al. (1985) and Collin-Souffrin et al. (1986)  found that the abundance of Fe in  QSOs, as measured from broad emission lines, turned out to be $\sim $ a factor of 10 more than the solar one and this represented a challenge for chemical evolution model makers.
Hamman \& Ferland (1992)  from N V/C IV line ratios in QSOs
derived the N/C abundance ratios and inferred the QSO metallicities. They  suggested that N is overabundant by  factors of 2-9 in the high redshift sources ($z >2$).
Metallicities 3-14 times the solar one were also suggested in order to produce
such a  high N abundance, under the assumption of a mainly secondary N.
To interpret their data they built a chemical evolution model, a Milky Way- like model, and
suggested that these high metallicities are reached in only 0.5 Gyr, implying that QSOs are associated with vigourous star formation. 
At the same time, Padovani \& Matteucci (1993) and Matteucci \& Padovani (1993)
proposed a model for QSOs in which QSOs are hosted by massive ellipticals.
They assumed that after the occurrence of a galactic wind the galaxy evolves passively and that for massess $> 10^{11} M_{\odot}$ the gas restored by the dying stars is not lost but it feeds the central black hole. 
They showed that in this context the stellar mass loss rate can explain the observed AGN luminosities.
They also found that solar abundances in the gas are reached in no more 
than $10^{8}$ years explaining in a natural way the standard emission lines 
observed in high-z QSOs.
The predicted abundances could explain the data available at that time and 
solve the problem of the quasi-similarity of QSO spectra at different 
redshifts. Finally, they suggested also a criterium for establishing the ages of 
QSOs on the basis of the [$\alpha$/Fe] ratios observed from broad emission 
lines (see also Hamman \& Ferland 1993). 

Much more recently,
Maiolino et al. (2005, 2006) used more than 5000 QSO spectra
from SDSS data to investigate the metallicity of the broad emission line  
region
in the redshift range $2 < z < 4.5$ and over the luminosity range 
$-24.5 < M_B < -29.5$.
They found substantial chemical enrichment in QSOs already at z = 6. Models for ellipticals by Pipino \& Matteucci (2004) were used as a comparison with the data and they well reproduce the data, as one can see in Figure 4.2. In this Figure the evolution of the abundances of several chemical elements in the gas of a typical elliptical are shown. The elliptical suffers a galactic wind at  around 0.4 Gyr since the beginning of star formation. This
wind devoids the galaxy of all the gas present at that time. After this time, 
the SF stops and the galaxy evolves passively. All the gas restored after the galactic wind event by dying stars can in principle feed the central black hole, thus the abundances shown in Figure 4.2, after the time of the wind, can be compared with the abundances measured in the broad emission line region. As one can see, the predicted Fe abundance after the galactic wind is always higher than the O one, owing to the Type Ia SNe which continue to produce Fe even after the stop in the SF. On the other hand, O and $\alpha$-elements stop to be produced when the SF halts. The comparison between the predicted abundances and those derived from the QSO spectra, are in very good agreement and indicates ages for these objects between 0.5 and 1 Gyr.

\begin{figure}
\includegraphics[width=4.5in,height=4.0in]{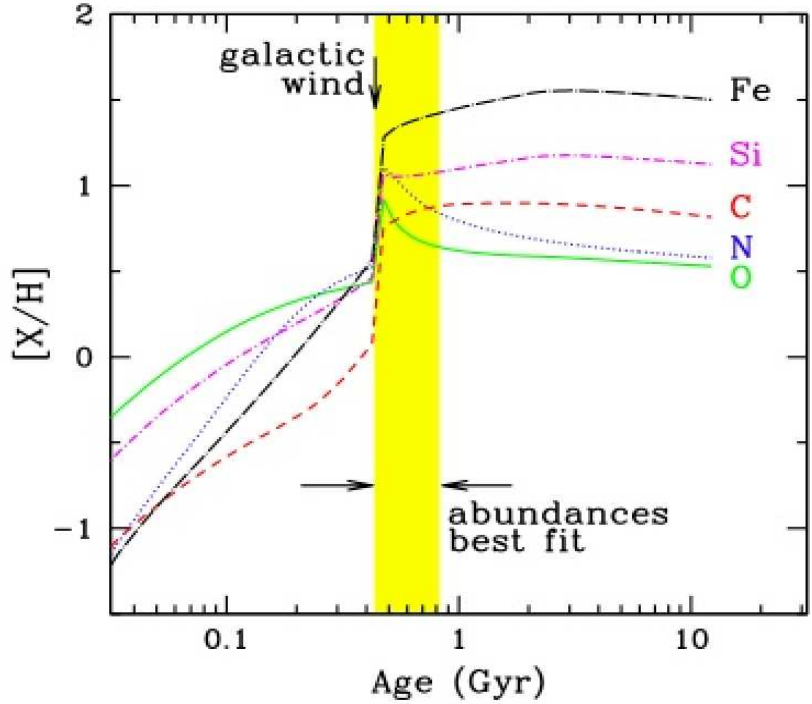}
\hfill
\caption{The temporal evolution of the abundances of several chemical elements in the gas of an elliptical galaxy with luminous mass of $10^{11}M_{\odot}$. Feedback effects are taken into account in the model (Pipino \& Matteucci 2004), as described in Lecture III. The downarrow indicates the time  for the occurrence of the galactic wind. After this time, the SF stops and the elliptical evolves passively. All the abundances after the time for the occurrence of the wind are those that we observe in the broad emission line region region. The shaded area indicates the abundance sets which best fit the line ratios observed in the QSO spectra. Figure from  Maiolino et al. 2006.}\label{fig} 
\end{figure}

Finally, in the context of the joint formation of QSOs and ellipticals we recall the work of Granato et al. (2001) who includes the energy feedback from the central AGN in ellipticals. This feedback produces outflows and stops the SF in a down-sizing fashion, in agreement with the chemical properties of ETGs indicating a shorter period of SF for the more massive objects.

\subsection{The chemical enrichment of the ICM}

The X-ray emission from galay clusters
is generally interpreted as thermal bremsstrahlung in a hot gas 
($10^{7}$-$10^{8}$ K).
There are several emission lines (O, Mg, Si, S)  including the strong Fe K-line at 
around 7keV which was discovered by Mitchell et al. (1976).
The iron is the best studied element in clusters. For $kT \ge$ 3 keV
the intracluster medium (ICM) Fe abundance is constant and $\sim 0.3 Fe_{\odot}$ in the central cluster 
regions; the existence of metallicity gradients seems evident only in some clusters (see Renzini 2004). 
At lower 
temperatures, the situation is not so simple and the Fe abundance seems to 
increase.
The first works on chemical enrichment of the ICM even preceeded the 
discovery of the Fe line (Gunn \& Gott 1972, Larson \&
Dinerstein 1975). In the following years other works appeared such as those of Vigroux (1977), Himmes \& Biermann (1988) and 
Matteucci \& Vettolani (1988). In particular, 
Matteucci \& Vettolani (1988) started a more 
detailed 
approach to the problem followed by David et al. (1991), Arnaud (1992), 
Renzini et al. (1993), Elbaz et al. (1995), Matteucci \& Gibson (1995), 
Gibson \& Matteucci (1997), Lowenstein  \& Mushotzky (1996), Martinelli 
et al. (2000), Chiosi (2000), Moretti et al. (2003).
The majority of these papers assumed that galactic winds  
(mainly from ellipticals and S0 galaxies) are responsible for the ICM chemical enrichment. In fact,
ETGs are the dominant type of galaxy in clusters and Arnaud (1992) found a clear correlation between the mass of Fe in clusters and the total luminosity of ellipticals. No such correlation was found for spirals in clusters.
Alternatively, the abundances in the ICM are due to ram 
pressure stripping (Himmes \& Biermann 1988) or derive from a chemical enrichment from pre-galactic 
Pop  III stars (White \& Rees 1978). 

In Matteucci \& Vettolani (1988) the Fe abundance in the ICM  relative 
to the Sun, $X_{Fe}/X_{Fe_{\odot}}$, was calculated as 
$(M_{Fe})_{pred}/(M_{gas})_{obs}$ to be compared with the observed ratio  $(X_{Fe}/X_{Fe_{\odot}})_{obs}=0.3-0.5$
(Rothenflug  \& Arnaud 1985). They found a good agreement with the observed Fe abundance in clusters if all the Fe produced by ellipticals and S0, after SF has stopped, is eventually restored into the ICM and if the majority of gas in clusters has a primordial origin.
Low values for [Mg/Fe] and [Si/Fe] were predicted  at the present time, 
due to the short period of SF  in ETGs and to the Fe produced 
by Type Ia SNe. With Salpeter IMF  they found that the Type Ia SNe contribute $ \ge 50\%$ of the total Fe in clusters. This leads to a bimodality in the [$\alpha$/Fe] ratios in the stars and in the gas in the ICM, since the stars have overabundances of [$\alpha$/Fe]$>$ 0 whereas the ICM should have 
[$\alpha$/Fe]$ \le 0$.
The same conclusion was reached  and more highlighted later by
Renzini et al. (1993). More recently, Pipino et al. (2002) 
computed the chemical 
enrichment of the ICM as a function of redshift by considering the evolution 
of the cluster luminosity function and an updated treatment of the SN feedback.
They adopted Woosley \& Weaver (1995) yields for Type II SNe and Nomoto et al. (1997) W7 model for Type Ia SNe and a Salpeter IMF. They also predicted solar or undersolar [$\alpha$/Fe] ratios in the ICM. The observational data on abundance ratios in clusters are still uncertain and vary from cluster centers where they tend to be solar or undersolar to the outer regions where they tend to be oversolar (e.g. Tamura et al. 2004). So, no firm  conclusions can be drawn on this point. Concerning the evolution of the Fe abundance in the ICM as a function of 
redshift,  most of the above mentioned models predict very little or no evolution of the Fe abundance from z=1 to z=0 (Pipino et al. 2002). This prediction seemed to be in good agreement with data from Tozzi et al. (2003) as shown Figure.
However, more recently, more data of Fe abundance for high redshift clusters appeared showing a different behaviour.
\begin{figure}
\includegraphics[width=4.5in,height=3.0in]{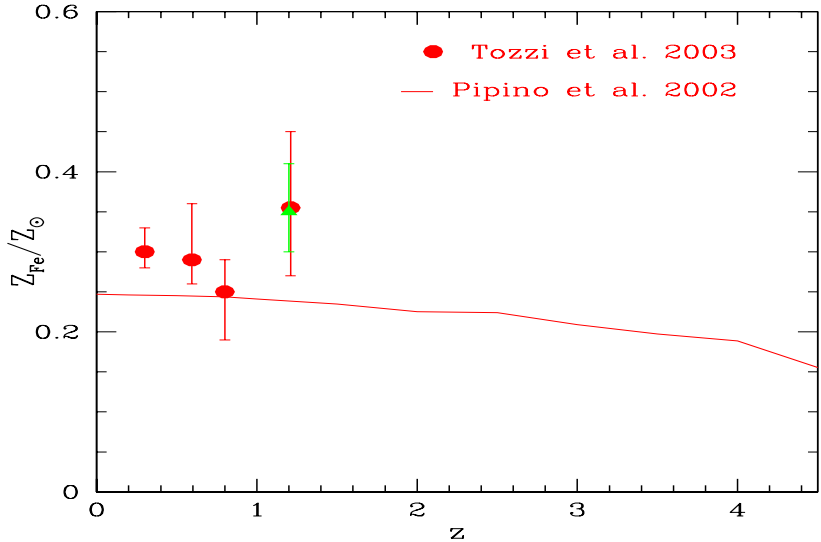}
\hfill
\caption{Observed Fe abundance and predicted Fe abundance in the ICM as a function of redshift: data from Tozzi et al. (2003), model (continuous line) from Pipino et al. (2002), where the formation of ETGs was assumed to occur at z=8.}\label{fig} 
\end{figure}

In Figure 4.4 we show the data of Balestra et al. (2006) who claim an increase, by at least a factor of two, of the Fe abundance in the ICM from z=1 to z=0. Clearly, if we assume that only ellipticals have contributed to the Fe abundance in the ICM, this effect is difficult to explain unless we assume recent star formation in ellipticals. Another possible explanation could be that spiral galaxies contribute to Fe when they become S0 as a consequence of ram pressure stripping, and this morphological transformation might have started just at z=1.

\begin{figure}
\includegraphics[width=4.5in,height=3.0in]{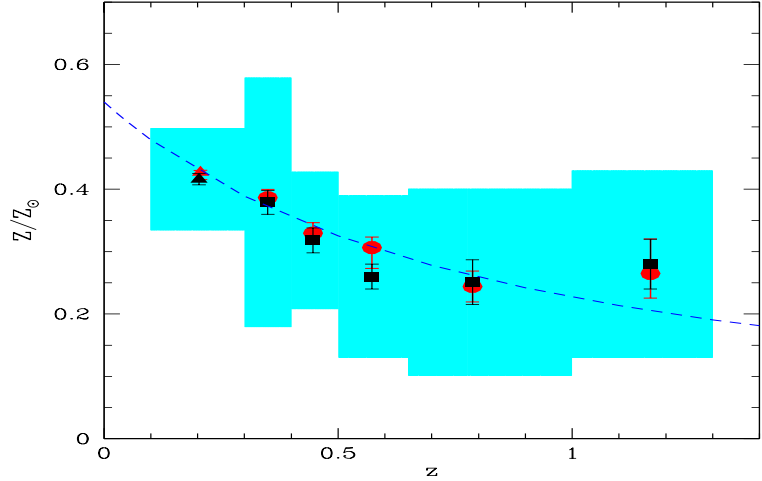}
\hfill
\caption{New data  (always relative to Fe) from Balestra et al. (2006) showing an increase of the Fe abundance in the ICM from z=1 to z=0. Error bars refer to 1$\sigma$ confidence level. The big shaded area represents the rms dispersion. Figure from Balestra et al. (2006). }\label{fig} 
\end{figure}

\subsection{Conclusions on the enrichment of the ICM}
From what said before we can conclude that:
\begin{itemize}

\item  Elliptical galaxies are the dominant contributors to the abundances and 
energetic content of the ICM.
A constant Fe abundance of $\sim 0.3 Fe_{\odot}$ is found in the central regions of clusters hotter 
than 3keV (Renzini 2004).

\item Good models for the chemical enrichment of the ICM should 
reproduce the iron mass measured in clusters plus the
[$\alpha$/Fe] ratios inside galaxies and in the ICM as well as the Fe mass to 
light ratio (IMLR= $M_{Fe_{ICM}}/L_{B}$, with $L_B$ being the total blue 
luminosity of member galaxies, as defined by Renzini et al. (1993). Abundance ratios 
are very powerful tools to impose constraints on the evolution of ellipticals and of the ICM.

\item 
Models which do not assume a top-heavy IMF for the galaxies in clusters 
(a Salpeter IMF can reproduce at best the properties of local ellipticals) 
predict
[$\alpha$/Fe]$>$ 0 inside ellipticals and [$\alpha$/Fe] $\le$ 0 in the ICM. 
Observed values are still too uncertain to draw firm conclusions on this point.

\end{itemize}

\vspace{20mm}
\noindent
{\bf Acknowledgements}

This research has been supported by INAF (Italian National Institute for Astrophysics), Project PRIN-INAF-2005-1.06.08.16

\end{document}